\documentclass[usenatbib]{emulateapj}
\usepackage{graphicx}
\usepackage[flushleft]{threeparttable}
\usepackage[usenames,dvipsnames,svgnames,table]{xcolor}
\usepackage{rotating}
\usepackage{hyperref}
\definecolor{darkblue}{rgb}{0.0,0.0,0.3}
\hypersetup{colorlinks,breaklinks,
            linkcolor=darkblue,urlcolor=darkblue,
            anchorcolor=darkblue,citecolor=darkblue}
\usepackage{amssymb}
\usepackage{amsmath}
\usepackage{tikz}
\usetikzlibrary{patterns}
\usepackage{array}
\usepackage[none]{hyphenat}

\makeatletter
\renewcommand{\p@subsection}{}
\renewcommand{\p@subsubsection}{}
\makeatother

\begin{document}

\def\etal{et al.\ \rm}
\def\ba{\begin{eqnarray}}
\def\ea{\end{eqnarray}}
\def\etal{et al.\ \rm}
\def\Fdw{F_{\rm dw}}
\def\Tex{T_{\rm ex}}
\def\Fdis{F_{\rm dw,dis}}
\def\Fnu{F_\nu}
\def\FJ{F_J}
\def\rout{r_{\rm out}}
\def\icarus{\rm Icarus}

\newcommand\cmtrr[1]{{\color{red}[RR: #1]}}
\newcommand\cmtid[1]{{\color{green}[ID: #1]}}

\newcommand{\pr}{\indent}
\newcommand{\no}{\noindent}
\newcommand{\be}{\begin{equation}}
\newcommand{\fr}{\frac}
\newcommand{\ee}{\end{equation}}
\newcommand{\pa}{\partial}
\renewcommand{\phi}{\varphi}
\newcommand{\ti}{\tilde}
\newcommand{\ep}{\varepsilon}


\title{Secular Evolution Driven by Massive Eccentric Disks/Rings: An Apsidally Aligned Case}

\author{Irina Davydenkova\altaffilmark{1} \& Roman R. Rafikov\altaffilmark{2,3}}
\altaffiltext{1}{Universit\'e de Gen\`eve, 2-4 rue du Li\`evre, c.p. 64, 1211 Gen\`eve~4, Switzerland; Irina.Davydenkova@unige.ch}
\altaffiltext{2}{Department of Applied Mathematics and Theoretical Physics, Centre for Mathematical Sciences, University of Cambridge, Wilberforce Road, Cambridge CB3 0WA, UK; rrr@damtp.cam.ac.uk}
\altaffiltext{3}{Institute for Advanced Study, 1 Einstein Drive, Princeton, NJ~08540, USA}


\begin{abstract}
Massive eccentric disks (gaseous or particulate) orbiting a dominant central mass appear in many astrophysical systems, including planetary rings, protoplanetary and accretion disks in binaries, and nuclear stellar disks around supermassive black holes in galactic centers. We present an analytical framework for treating the nearly Keplerian secular dynamics of test particles driven by the gravity of an eccentric, apsidally aligned, zero-thickness disk with arbitrary surface density and eccentricity profiles. We derive a disturbing function describing the secular evolution of coplanar objects, which is explicitly related (via one-dimensional, convergent integrals) to the disk surface density and eccentricity profiles without using any ad hoc softening of the potential. Our analytical framework is verified via direct orbit integrations, which show it to be accurate in the low-eccentricity limit for a variety of disk models (for disk eccentricity $\lesssim 0.1-0.2$). We find that free precession in the potential of a disk with a smooth surface density distribution can naturally change from prograde to retrograde within the disk. Sharp disk features --- edges and gaps --- are the locations where this tendency is naturally enhanced, while the precession becomes very fast. Radii where free precession changes sign are the locations where substantial (formally singular) growth of the forced eccentricity of the orbiting objects occurs. Based on our results, we formulate a self-consistent analytical framework for computing an eccentricity profile for an aligned, eccentric disk (with a prescribed surface density profile) capable of precessing as a solid body under its own self-gravity.
\end{abstract}



\section{Introduction}  
\label{sect:intro}


Astrophysical disks orbiting in the gravitational potential of a dominant central mass $M_c$ often possess nonaxisymmetric shape. The nonaxisymmetric distortion can be modeled as a manifestation of {\it disk eccentricity}. In this picture, to zeroth order, different components of the disk --- parcels of gas in fluid (collisional) disks, or particles (e.g. stars) in collisionless disks  --- move on eccentric Keplerian orbits in the field of a central mass. Even if the mass of the disk $M_d$ is much smaller than $M_c$, the self-gravity of the disk can still play a very important role in its dynamics, as well as the orbital evolution of external objects, by driving precession and causing an exchange of angular momentum between different parts of the disk on long (secular) timescales. 

Such eccentric disks are encountered in a variety of astrophysical contexts --- galactic, stellar, and planetary. One of the closest examples is provided by the eccentric planetary rings, such as the $\epsilon$, $\alpha$, and $\beta$ rings\footnote{The latter two are also significantly inclined with respect to the equatorial plane of the planet.} of Uranus \citep{Elliot}, as well as the Titan and Maxwell ringlets of Saturn \citep{Porco}. These particulate collisional rings are very narrow, essentially representing limiting cases of eccentric disks with the spread in semimajor axes of their constituent particles $\Delta a\lesssim e_r a_r$, where $a_r$ and $e_r$ are the mean semimajor axis and eccentricity of the rings. 
As demonstrated by \citet{GT}, self-gravity of the rings, coupled with collisional effects \citep{CG2000,CC2003,Mosqueira,Pan}, can counter differential precession driven by the planetary oblateness, allowing rings to precess as a solid body while maintaining a coherent eccentric shape.

A significant number of stellar binaries are known to host exoplanets, orbiting either the whole system \citep{Doyle,Welsh} or one of the binary components \citep{Hatzes,Chauvin}. The formation and early dynamics of such planets are significantly complicated by the fact that the nonaxisymmetric binary potential excites the nonzero eccentricity of the protoplanetary disks \citep{Kley,Regaly,Miranda}, in which the building blocks of these planets --- planetesimals --- orbit. It has been recently shown  \citep{Rafikov13a,Rafikov13b,SR15a,SR15b,RS15a,RS15b} that the gravitational effect of such an eccentric protoplanetary disk plays a key role in planetesimal dynamics for young binaries.

Spectroscopic observations of accretion disks in cataclysmic variables using the technique of Doppler tomography \citep{Marsh} suggest that a certain type of variability in these systems --- the so-called ``superhump" \citep{Horne} --- is caused by the precession of an eccentric accretion disk around the white dwarf \citep{Lubow}. Asymmetric evolving emission lines found in the spectra of compact disks of gaseous debris around some metal-rich white dwarfs \citep{Gansicke} also provide evidence for their nonzero eccentricity \citep{D18,Miranda2018}.  

Finally, optical emission from the central region of the M31 galaxy exhibits a double-nucleus morphology \citep{Bacon,Bender}. The best interpretation of the existing photometric and spectroscopic data, first proposed by \citet{Tremaine95}, points at the existence of a highly eccentric ($e_d\sim 0.5$) stellar disk orbiting the central supermassive black hole. A number of models, both purely kinematic, i.e. not accounting for the disk self-gravity in maintaining its coherence \citep{Peiris,Brown}, and as fully dynamic \citep{Salow2001,Bacon,Sambhus,Salow}, have been put forward to understand  this system. Our own Galaxy harbors an eccentric disk of young stars orbiting the central supermassive black hole \citep{Levin,Bartko}, whose gravity may affect its own dynamics \citep{Nayakshin}. 

In many of these systems, disk gravity plays the dominant role in disk dynamics, as well as in the orbital evolution of nearby objects (e.g. planetesimals in protoplanetary disks in binaries). This motivated a number of past analytical \citep{GT,CG2000} and numerical \citep{Bacon,Sambhus,Salow,Nayakshin} studies aimed at clarifying the details of the dynamics driven by the gravity of an eccentric disk. Such calculations inevitably require an efficient method for computing the potential $\Phi_d$ of an eccentric disk at every point. Moreover, since $M_d\ll M_c$, the disk-driven evolution is typically rather slow, justifying the use of a {\it secular} approximation, in which the disk potential is averaged over the orbital motion of an object under consideration. A direct calculation of such an averaged potential, or a {\it secular disturbing function} as it is known in celestial mechanics, in general requires evaluation of three-dimensional integrals (see equation (\ref{mainint})), which is impractical in many applications.  

\citet{SR15a} presented a calculation of a secular disturbing function for a particular model of a {\it radially extended} (i.e. having $\Delta a\sim a$), apsidally aligned  eccentric disk. They assumed that both the surface density and the eccentricity of the disk vary as {\it power laws} of the semimajor axis $a$ of the mass elements comprising the disk. Their resultant disturbing function does not involve multidimensional integration and can be used for efficient analysis of disk-driven orbital dynamics. In particular, it was employed to provide a self-consistent treatment of the secular evolution of planetesimals orbiting in a massive eccentric protoplanetary disk within (or around) a young stellar binary  \citep{RS15a,RS15b,SR15b}. 

The goal of our present work is to provide a natural but important generalization of the results of \citet{SR15a}. Here we develop a general analytical framework for computing a secular disturbing function for an apsidally aligned\footnote{The assumption of apsidal alignment is retained here for simplicity. It is relaxed in a subsequent work of Davydenkova \& Rafikov (in prep.).}, eccentric disk with {\it arbitrary} radial profiles of the disk surface density $\Sigma$ and eccentricity $e_d$. We show that this disturbing function can be reduced to a combination of one-dimensional integrals over the radial profiles of $\Sigma$ and $e_d$, enabling application of our results to a broad range of practical problems (e.g. computation of the structure of a rigidly precessing eccentric disk, see \S \ref{sect:disc}). We also provide numerical verification of our analytical results using direct orbit integrations.

Our work is organized as follows. We describe our methodology and outline the results of the disturbing function calculation in \S \ref{sect:method}; the details of its derivation can be found in Appendix \ref{sect:disk_potential}. We describe the strategy for numerical verification of our analytical calculations in \S \ref{sect:numerics} and then present our findings in \S \ref{sect:results}. Having tested our analytical framework, we then describe several of its applications in \S \ref{sect:apply}, including the derivation of a self-consistent method for calculating the eccentricity distribution of an eccentric, apsidally aligned disk (with a prescribed surface density distribution) that can precess as a solid body while maintaining its overall shape (\S \ref{sect:modes}). We discuss our results in \S \ref{sect:disc} and provide a brief summary in \S \ref{sect:sum}.


\section{Secular disturbing function}  
\label{sect:method}


Our goal is to calculate the secular (i.e. orbit-averaged) gravitational potential felt by a test particle orbiting in the combined gravitational field of a central mass $M_d$ and an eccentric disk (fluid or particulate). This particle can be an external object or it can be one of the mass elements comprising the disk (see \S \ref{sect:modes}). The test particle moves on an eccentric orbit coplanar with the disk, with semimajor axis $a_p$, eccentricity $e_p$, and apsidal angle $\varpi_p$.  

\begin{figure}
\centering
\includegraphics[width=0.5\textwidth]{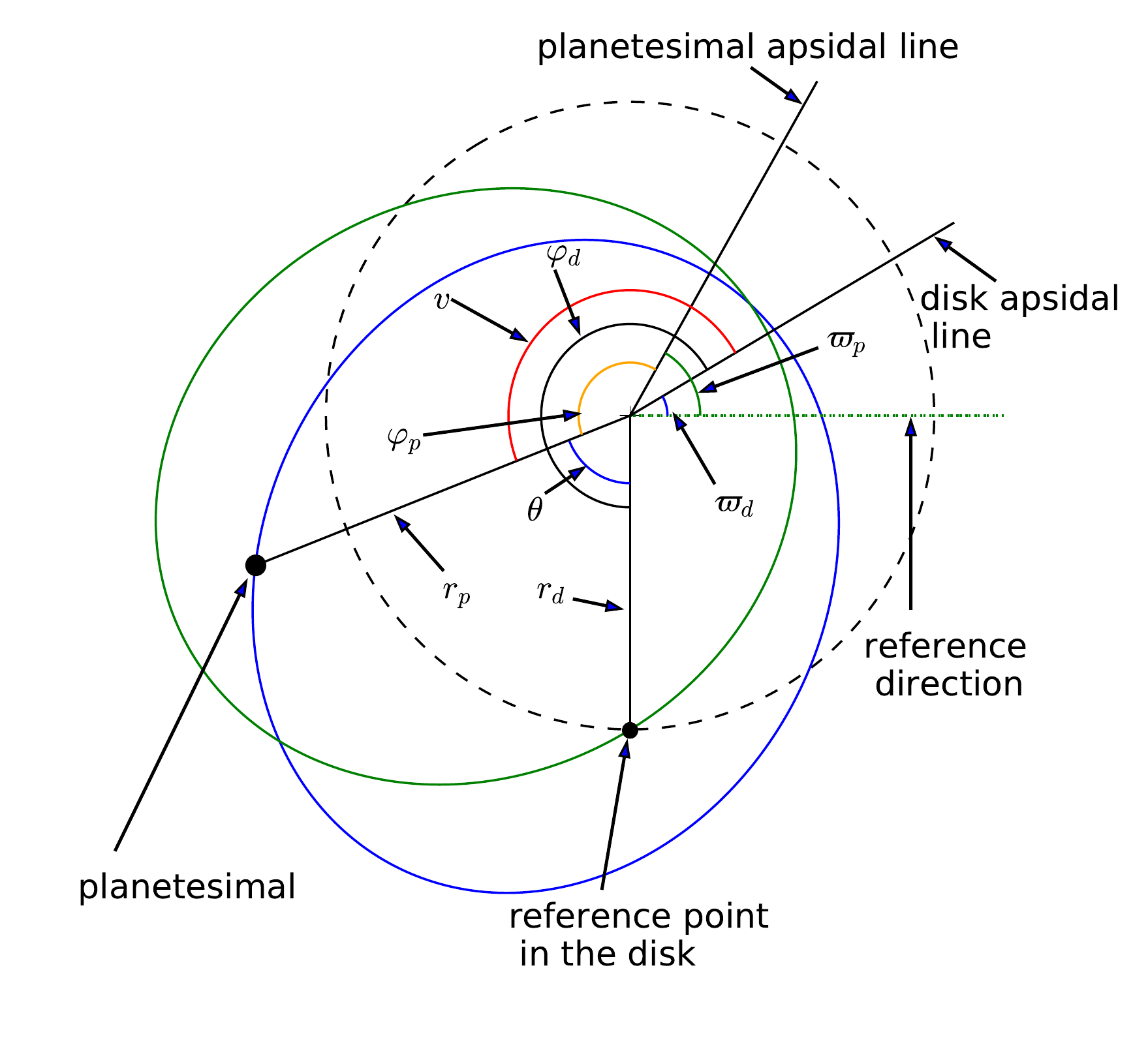}
\caption{Geometry of the problem, showing elliptical trajectories of the test particle (blue) and a mass element of the disk (green). See text for details.
\label{fig:geometry}}
\end{figure}

The disk is purely two-dimensional (i.e. it has zero vertical thickness) and is not warped (i.e. lies in a single plane). It is eccentric and apsidally aligned in the sense that the trajectories of its constituent mass elements (fluid or particle) are {\it confocal Keplerian ellipses}, which are {\it apsidally-aligned} in the direction making an angle $\varpi_d$ with respect to the reference direction. We define $r_d$ to be the distance from the common focus of the eccentric orbits of the constituent particles and $\phi_d$ to be the polar angle with respect to the disk apsidal line; see Figure \ref{fig:geometry} for illustration. 

For every trajectory with a semimajor axis $a$, we can define the disk {\it surface density at the periastron} $\Sigma_d(a)$ and the eccentricity of the fluid trajectory $e_d(a)$, which we will simply call {\it disk eccentricity}. The disk mass distribution can also be characterized by the mass per unit semimajor axis $\mu(a)$. Using the basic properties of Keplerian dynamics, one can show that $\Sigma_d$ and $\mu$ are directly related via
\ba   
\mu(a)=2\pi a\Sigma_d(a)\left[\frac{1+e_d(a)}{1-e_d(a)}\right]^{1/2}\left[1-e_d(a)(1+\zeta)\right]
\label{eq:mu},
\ea   
where $\zeta \equiv d\ln e_d(a)/d\ln a$. In general, $\Sigma_d(a)$ (or $\mu(a)$) and $e_d(a)$ can be arbitrary functions of the semimajor axis $a_d$, as long as $e_d(a)$ varies slowly enough for the particle trajectories to be noncrossing\footnote{This requires $de_d/d\ln a<(1-e_d)$ \citep{Ogilvie,Statler2001}.}. In this work, we choose $\Sigma_d$ (rather than $\mu$) to characterize the distribution of mass in the disk.

Note that in the secular approximation relied upon in this study, the energy and semimajor axes of particles (or fluid elements) comprising the disk are the intergals of motion. As a result, the amount of disk mass per unit semi-major axis $\mu(a)$ is strictly conserved even if the disk shape changes. Consequently, according to equation (\ref{eq:mu}), if $e_d$ does not change in time, then $\Sigma_d(a)$ is also independent of time (this will be important in \S \ref{sect:modes}). 

\citet{Statler2001} and \citet{Ogilvie} provided an expression for the two-dimensional surface density $\Sigma(r_d,\phi_d)$ of an eccentric disk in terms of $e_d(a)$ and disk mass distribution $\mu(a)$. For our present purposes, it is more convenient to write $\Sigma$ as a function of $a$ and $\phi_d$, relating it to $\Sigma_d(a)$. Using the calculation of $\Sigma(a,\phi_d)$ in \citet{Statler1999}, for an apsidally aligned disk, we can write
\ba
\Sigma(a,\phi_d)=\Sigma_d(a)\frac{1-e_d^2-\zeta e_d (1+e_d)}
{1-e_d^2-\zeta e_d \left[e_d+\cos E(\phi_d)\right]},
\label{eq:Sigma}
\ea
where $E(\phi_d)$ is the eccentric anomaly ($E=\phi_d=0$ at pericenter) and $\Sigma_d$ and $e_d$ are functions of $a$. 

Even though the expression (\ref{eq:Sigma}) holds for arbitrary $e_d$, in the rest of the paper, we will take the eccentricities of both the disk and test particle to be small, $e_d(r)\ll 1$ and $e_p\ll 1$. This is needed for our secular theory (formulated at the lowest order in eccentricity) to provide an accurate description of orbital dynamics. As a consequence of this approximation, equation (\ref{eq:mu}) also yields  $\Sigma_d(a)\approx \mu(a)/(2\pi a)$ to lowest order in $e_d$. Thus, even if $e_d\ll 1$ varies in time, $\Sigma_d(a)$ should still be conserved to $O(e_d)$ accuracy in the course of secular evolution.


\subsection{Secular (Orbit-averaged) potential of the disk}
\label{sect:sec}


Our calculation of the orbit-averaged disturbing function $R_d$ due to an eccentric disk uses the general mathematical procedure outlined in a seminal paper of \citet{Heppen} for the calculation of the disturbing function due to an {\it axisymmetric} disk. In this approach, the expansion of the disturbing function in terms of a small parameter --- test particle eccentricity --- proceeds differently from the classical Laplace-Lagrange theory \citep{MurrayDermott}. The resultant expression for $R_d$ does not contain non-integrable singularities at the particle semimajor axis, resulting in a convergent expression for the disturbing function. In other words, this calculation does not require introduction of an ad hoc {\it softening} of the potential. This method was later used by \citet{Ward} to study the stability of the early Solar System perturbed by an axisymmetric protoplanetary disk.

\citet{SR15a} extended the method of \citet{Heppen} to the case of {\it non-axisymmetric, apsidally aligned, eccentric} disks with $\Sigma$ given by equation (\ref{eq:Sigma}). However, their work was restricted to disks with power-law profiles of $\Sigma_d(a)$ and $e_d(a)$. Here we generalize this calculation even further to cover the disks with {\it arbitrary} behaviors of $\Sigma_d(a)$ and $e_d(a)$. 

As a result of a rather lengthy derivation, the mathematical details of which are presented in Appendix \ref{sect:disk_potential}, we arrive at the following expression for the disturbing function due to an apsidally aligned disk: 
\ba
R_d = a_p^2 n_p\left[\frac{1}{2}A_d e_p^2+B_d e_p\cos(\varpi_p-\varpi_d)\right],
\label{eq:R_gen}
\ea
where $n_p=(GM_c/a_p^3)^{1/2}$ is the particle mean motion, with the coefficients $A_d$ and $B_d$ (having dimensions of [s$^{-1}$] and discussed in more detail in \S \ref{sect:AdBd}) given by the expressions (\ref{eq:AdBd})-(\ref{eq:Bd_edge}). The key underlying simplifications making this calculation possible are that $e_d\ll 1$, as well as $d e_d/d\ln a\ll 1$; see equation (\ref{eq:Sig_expand}). 

Introducing a two-component eccentricity vector ${\bf e}_p=(k_p,h_p)=e_p(\cos\varpi_p,\sin\varpi_p)$ for a test particle, as well as the auxiliary vector 
\ba   
{\bf B}_d=B_d(\cos\varpi_d,\sin\varpi_d),
\label{eq:e_d}
\ea   
the expression (\ref{eq:R_gen}) can be rewritten as
\ba  
&& R_d = a_p^2 n_p\left[\frac{1}{2}A_d {\bf e}_p^2+{\bf e}_p\cdot {\bf B}_d\right]
\nonumber\\
&& =a_p^2 n_p\left[\frac{A_d}{2}\left(k_p^2+h_p^2\right)+B_d\left(k_p\cos\varpi_d + h_p\sin\varpi_d\right)\right].
\label{eq:R1}
\ea  
Note that ${\bf B}_d$ is different from the disk eccentricity vector ${\bf e}_d=e_d(\cos\varpi_d,\sin\varpi_d)$; in fact, ${\bf B}_d$ involves a convolution of $e_d(a)$ with a complicated kernel (see equation (\ref{eq:Bd_bulk})) over the radial extent of the disk. 

The mathematical structure of $R_d$ in equation (\ref{eq:R_gen}) is identical to that of the conventional Laplace-Lagrange secular disturbing function \citep{MurrayDermott}. The difference between them lies in the explicit dependence of the coefficients $A_d$ and $B_d$ on $a_p$. The behavior of  $A_d(a_p)$ and $B_d(a_p)$ is determined, eventually, by the radial profiles of the disk surface density $\Sigma_d$ and eccentricity $e_d$. The exploration of this behavior is the major focus of our study.


\subsection{Mathematical properties of $A_d$ and $B_d$}
\label{sect:AdBd}

The expression (\ref{eq:AdBd}) for $A_d$ consists of two parts. One of them, $A_d^{\rm bulk}$, involves integral convolution over the disk semimajor axis $a$ of a (prescribed) disk surface density $\Sigma_d(a)$, as well as its radial derivatives up to second order, all of which enter linearly; see equation (\ref{eq:Ad_bulk}). The convolution kernel involves a Laplace coefficient $b^{(0)}_{1/2}(\alpha)$, which features a weak (logarithmic) singularity arising as $\alpha=a/a_p\to 1$ (i.e. for disk annuli close to the particle orbit). This singularity is, however, fully integrable and vanishes upon integration over the radial extent of the disk (see Appendix \ref{sect:disk_potential}). As always, the  Laplace coefficients are defined as
\ba   
b^{(j)}_{s}(\alpha)=\pi^{-1}\int_0^{2\pi}\frac{\cos( j\psi) d\psi}{\left(1-2\alpha\cos\psi+\alpha^2\right)^s}.
\label{eq:Laplace}
\ea   
    
In addition, whenever the disk has sharp edges (or discontinuous transitions in surface density), the expression (\ref{eq:AdBd}) for $A_d$ features {\it boundary terms} $A_d^{\rm edge}$, given by the equation (\ref{eq:Ad_edge}). These terms involve $b^{(0)}_{1/2}(\alpha)$ and its derivative $b^{(0)\prime}_{1/2}(\alpha)$, both evaluated at $\alpha=a_p/a_e$, where $a_e$ is the semimajor axis of the disk edge. As the semimajor axis of a test particle $a_p$ approaches the disk edge, both $b^{(0)}_{1/2}(\alpha)$ and $b^{(0)\prime}_{1/2}(\alpha)$ diverge: $b^{(0)}_{1/2}(\alpha)\sim\ln|1-\alpha|$, while $b^{(0)\prime}_{1/2}(\alpha)\sim |1-\alpha|^{-1}$, where in this case $\alpha=a_e/a_p\to 1$. As a result, boundary terms, as well as $A_d$, diverge at the sharp disk edge. This singularity of the secular disturbing function is further explored in \S \ref{sect:results}.

The divergence of $A_d$ at the disk edge does not arise if $\Sigma_d(a)$ goes to zero at the boundary in a sufficiently smooth fashion. Indeed, $b^{(0)\prime}_{1/2}(\alpha)$ in the boundary terms in equation (\ref{eq:Ad_edge}) is multiplied by $\Sigma_d$ at the edge; $b^{(0)}_{1/2}(\alpha)$ is multiplied by both $\Sigma_d$ and its radial derivative $\Sigma_d^\prime$. As a result, boundary terms vanish whenever $\Sigma_d\propto |a-a_e|^{\kappa}$ and $\kappa>1$ near the edge at $a_e$. Thus, in disks with $\Sigma_d$ smoothly (faster than linearly in $|a-a_e|$) turning to zero at a finite semimajor axis $a_e$ the coefficient $A_d$ has no boundary contributions and, hence, does not diverge at the boundary. The same is also true for disks without edges, which have their surface density smoothly declining to zero as $a\to 0$ and $a\to \infty$. Both possibilities will be explored further in \S \ref{sect:results}.

Similarly, the expression (\ref{eq:AdBd}) for $B_d$ involves radial integrals over the product of $\Sigma_d(a)$ and the disk eccentricity $e_d(a)$, as well as their derivatives up to second order (contribution $B_d^{\rm bulk}$ given by equation (\ref{eq:Bd_bulk})), in addition to the boundary terms $B_d^{\rm edge}$ represented by equation (\ref{eq:Bd_bulk}). Once again, $B_d^{\rm edge}$ vanishes whenever $\Sigma_d$ goes to zero at the disk edges sufficiently rapidly, e.g. as $\Sigma_d\sim |a-a_e|^{\kappa}$ with $\kappa>1$. 

Other features of $A_d(a_p)$ and $B_d(a_p)$ behavior and their effect on secular dynamics will be discussed in \S \ref{sect:results}-\ref{sect:disc}.


\section{Numerical verification}
\label{sect:numerics}


Having derived the analytical framework embodied in equations (\ref{eq:R_gen}) and (\ref{eq:AdBd})-(\ref{eq:Bd_edge}), we also provide its numerical verification. We do this by comparing the eccentricity evolution of test particles computed based on our analytical results with the results of direct orbit integration in the potential of an eccentric disk (computed numerically). Details of both approaches, as well as the disk models used in this comparison, are outlined below.


\subsection{(Semi-)analytical orbital evolution}
\label{sect:semi-an}

One approach to calculating orbital evolution in the potential of an eccentric disk uses Lagrange equations for the evolution of orbital elements, $\dot k_p= - (a_p^2 n_p)^{-1}\partial R_d/\partial h_p$, $\dot h_p = (a_p^2 n_p)^{-1}\partial R_d/\partial k_p$, where we use the secular expression (\ref{eq:R_gen}) for the disturbing function $R_d$. As a result one, finds 
\ba
\dot k_p= - A_d h_p - B_d\sin\varpi_d,~~\dot h_p=A_d k_p+B_d\cos\varpi_d,
\label{eq:sec_eq}
\ea
with the general solution given by the superposition of the free and forced eccentricity vectors \citep{MurrayDermott}:
\ba
{\bf e}_p &=& {\bf e}_{\rm free}(t)+{\bf e}_{\rm forced},
\label{eq:gen_sol}\\
{\bf e}_{\rm free}(t) &=& e_{\rm free}\left(\cos\left(A_dt+\varpi_0\right),\sin\left(A_dt+\varpi_0\right)\right),
\label{eq:efree}\\
{\bf e}_{\rm forced} &=& e_{\rm forced}\left(\cos\varpi_d,\sin\varpi_d\right).
\label{eq:eforced}
\ea
Here the forced eccentricity is
\ba
e_{\rm forced}(a) &=& -\frac{B_d(a)}{A_d(a)},
\label{eq:e_forced}
\ea
and constants $e_{\rm free}>0$ and $\varpi_0$ are such that at $t=0$, equation (\ref{eq:gen_sol}) satisfies the initial condition ${\bf e}_p(0) = (k(0),h(0)) = e_p(0)(\cos\varpi_p(0),\sin\varpi_p(0))$, where $e_p(0)$ and $\varpi_p(0)$ are the initial eccentricity and periastron angle of an object.

In particular, an object starting on a circular orbit ($e_p(0)=0$) has $e_{\rm free}=e_{\rm forced}$, and its motion is described by 
\ba
e_p(t) &=& \left|\frac{2B_d}{A_d}\sin\frac{A_d t}{2}\right|,
\label{eq:ecc_circ}\\
\tan\varpi_p(t) &=& \tan\left(\frac{A_d t}{2}+\varpi_d+\frac{\pi}{2}\right),
\label{eq:varp_circ}
\nonumber
\ea
where $\varpi_p$ stays in the range $(0,2\pi)$. We will use this solution for numerical verification of our semi-analytical results, although any alternative solution corresponding to different initial conditions would work just as well. 

Full determination of ${\bf e}_p$ requires calculating $A_d$ and $B_d$ for the given profiles of the disk surface density $\Sigma_d(a)$ and eccentricity $\varpi_d(a)$. We do this with the help of equations (\ref{eq:AdBd})-(\ref{eq:Bd_edge}) by numerically evaluating the corresponding integrals. Despite the weak (logarithmic) singularity of the integrands, the integrals themselves are convergent and are calculated directly without resorting to any type of softening of the integrand near its singular point (which is often done in studies of disk dynamics; see, e.g. \citep{Tremaine2001,Hahn,Touma}). The lack of an ad hoc parameter (softening length) in our calculation is its important distinctive feature.

Once $A_d$ and $B_d$ are calculated, equations (\ref{eq:ecc_circ}) (or, more generally, equations (\ref{eq:gen_sol})-(\ref{eq:e_forced})) provide a full semi-analytical description of the particle motion in the disk potential.


\subsection{Direct orbit integration}
\label{sect:direct}

An alternative method to compute particle motion uses the direct orbit integrator MERCURY \citep{Chambers}, which employs the Bulirsch-Stoer algorithm \citep{NR}. This integration uses as its input the gravitational acceleration ${\bf g}$ due to the disk potential, which is calculated on a grid of $(245,100)$ points\footnote{We also tried denser grids but have not found any difference in the outcomes.} in $(r,\phi)$. Values of ${\bf g}$ on the grid are then interpolated to provide accurate accelerations everywhere in the disk.

At each grid point, ${\bf g}$ is computed by direct summation of $\nabla \Phi_d$ (see equation (\ref{mainint})) over the disk surface, with the surface density given by equation (\ref{eq:Sigma}) in its {\it exact} form, i.e. not performing small-$e_d$ expansion. This 2-dimensional integral is convergent in the Cauchy principal value sense, even though the integrand diverges at the point where acceleration is calculated. To avoid this mathematical singularity, our numerical evaluation is performed at a small height ($10^{-3}$ AU or smaller) above the disk; we make sure that the result is convergent with respect to the value of this height. Note that this procedure is not equivalent to the introduction of softening in our theoretical calculation.

We integrate particle orbits starting with $e_p=0$. This allows us to directly compare the results with the solution (\ref{eq:ecc_circ}) following from our semi-analytical calculation, thus directly verifying the accuracy of our framework for the secular evolution in the potential of an eccentric disk.


\subsection{Summary of the disk models used}
\label{sect:disk_models}

In our comparison effort, we use the following model distributions of the disk eccentricity $e_d(a)$:
\ba 
e_p(r) &=& e_0\left(1+\fr{a_2-a}{a_2-a_1}\right), 
\label{eq:LinEcc}\\
e_p(r) &=& e_0\frac{(a-a_1)^2(a-a_2)^2}{(a_2-a_1)^4},
\label{eq:SqEcc}
\ea
and surface density at periastron $\Sigma_d(a)$, 
\ba 
\Sigma_d(a) &=& \Sigma_0\frac{(a-a_1)^2(a-a_2)^2}{(a_2-a_1)^4},  \label{eq:QuadSigma}\\
\Sigma_d(a) &=& \Sigma_0 \exp\left[\frac{4-\left[(a/a_c) + (a_c/a)\right]^2}{0.18}\right], 
\label{eq:GaussSigma}\\
\Sigma_d(a) &=& \Sigma_0\left[1 + 4\sin\left(\pi\fr{a- a_1}{a_2 - a_1}\right)\right]. 
\label{eq:SinSigma}
\ea
with $a_1=0.1$ AU, $a_2=5$ AU, $a_c=1.5$ AU and $e_0$ and $\Sigma_0$ being the normalization factors that we vary in our models. In our calculations, we always assume the mass of the central star to be 1 $M_\odot$.

\begin{figure}
\centering
\includegraphics[width=0.5\textwidth]{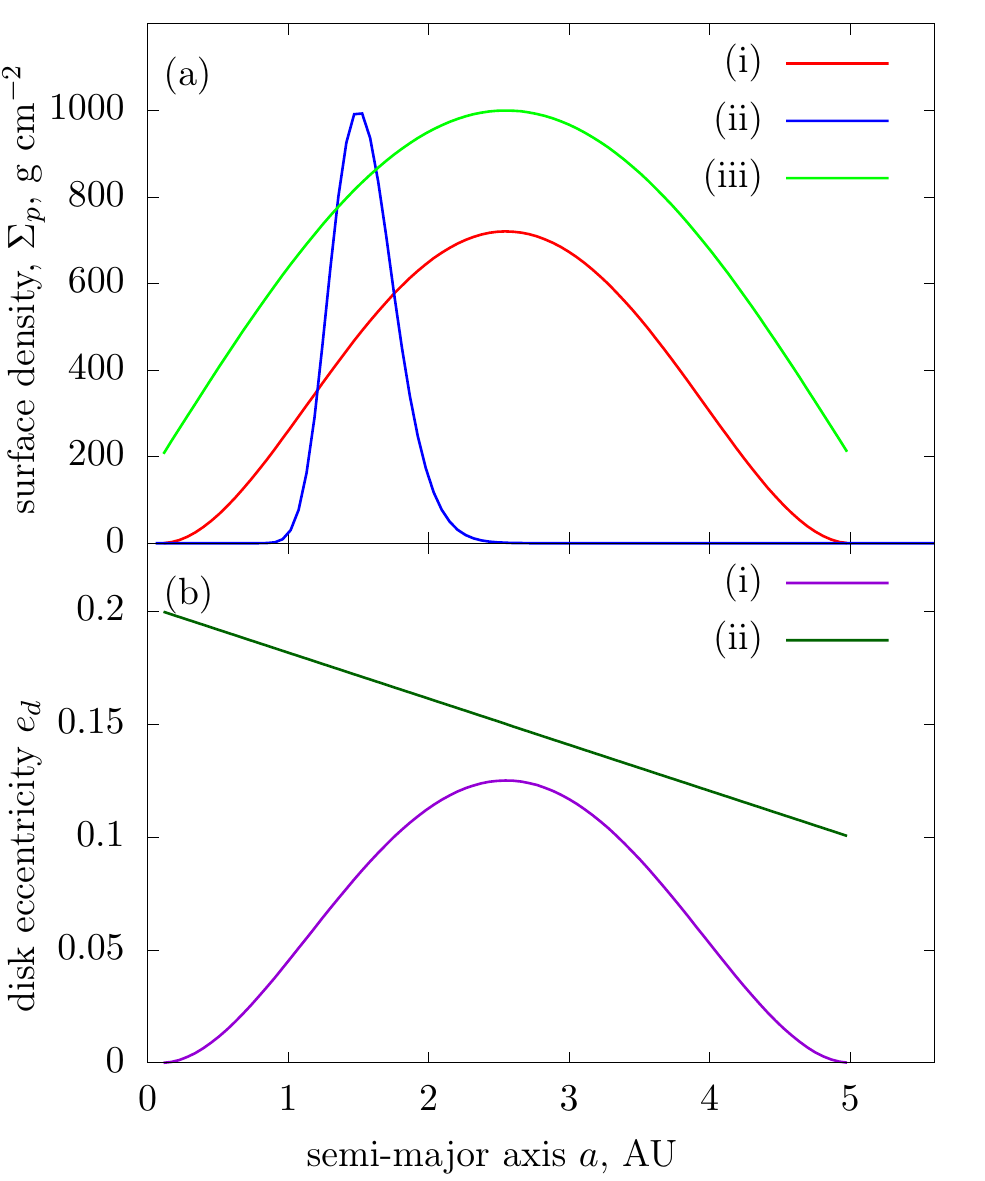}
\caption{
Profiles of the disk surface density $\Sigma_d(a)$ and eccentricity $e_d(a)$ considered in this work. Panel (a) shows the following $\Sigma_d(a)$ profiles: (i) is given by formula (\ref{eq:QuadSigma}) with $\Sigma_0\approx 1100$ g cm$^{-2}$, (ii) - by formula (\ref{eq:GaussSigma}) with $\Sigma_0=100$ g cm$^{-2}$, (iii) - by formula (\ref{eq:SinSigma}) with $\Sigma_0=20$ g cm$^{-2}$. 
Panel (b) shows the $e_d(a)$ profiles as follows: (i) is given by formula (\ref{eq:SqEcc}) with $e_0=2$ and (ii) by formula (\ref{eq:LinEcc}) with $e_0=0.1$.
\label{fig:SigmaProfile}}
\end{figure}

\begin{figure*}
\centering
\includegraphics[width=01.0\textwidth]{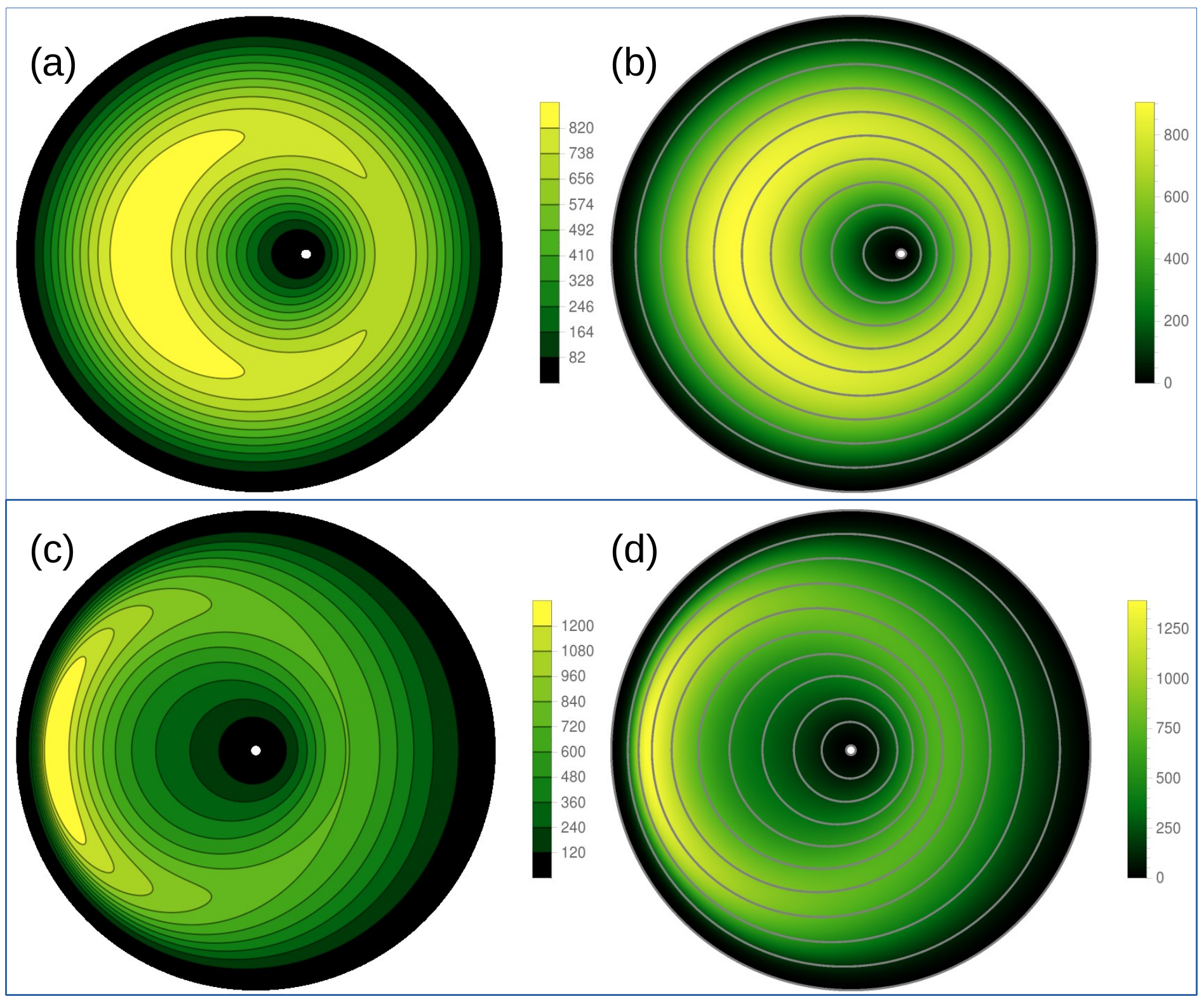}
\caption{
Maps of the disk surface density $\Sigma(r,\phi)$ (color indicates the amplitude of $\Sigma$) with overlaid contours of $\Sigma$ (thin curves; left panels) and eccentric trajectories of the mass elements comprising the disk (grey; right panels). The disk is always oriented with the apsidal lines pointing to the right. Top panels (a,b) are drawn for $\Sigma_d(a)$, given by equation (\ref{eq:QuadSigma}) with $\Sigma_0\approx 1100$ g cm$^{-2}$, and $e_d(a)$ given by (\ref{eq:LinEcc}) with $e_0=0.2$ (this high $e_0$ was chosen to better illustrate the $\Sigma$ distribution). Bottom panels (c,d) are drawn for the same $\Sigma_d(a)$ but a different disk eccentricity profile (\ref{eq:SqEcc}) with $e_0=0.5$. Note the substantial difference in $\Sigma(r,\phi)$ caused by just the difference in the $e_d(a)$ profiles.
\label{fig:green_maps}}
\end{figure*}

Figure \ref{fig:SigmaProfile} illustrates the behavior of these profiles of $\Sigma_d$ and $e_d$. In the majority of our calculations, we use the linear eccentricity profile (\ref{eq:LinEcc}), corresponding to a disk that is everywhere eccentric. Disks with the eccentricity profile (\ref{eq:SqEcc}), studied in \S \ref{sect:ecc_prof}, have circular inner and outer edges but are eccentric in between. 

The surface density profile (\ref{eq:QuadSigma}) corresponds to a disk in which $\Sigma_d$ smoothly goes to zero at the edges at $a_1$ and $a_2$. Based on the discussion in \S \ref{sect:AdBd}, we expect $A_d$ and $B_d$ to not diverge at the edges of such a disk, which is verified in \S \ref{sect:quad}. Profile (\ref{eq:GaussSigma}) describes a disk without edges (\S \ref{sect:Gauss}), which has its surface density exponentially decreasing for both $a\ll a_c$ and $a\gg a_c$. This disk also does not feature boundary terms (as it has no boundaries). Finally, the $\Sigma_d$ profile (\ref{eq:SinSigma}) describes a disk with discontinuous drops of the surface density at the edges. In this disk, we expect $A_d$ and $B_d$ to diverge at the boundaries, which is verified in \S \ref{sect:sharp}.

Figure \ref{fig:green_maps} illustrates the 2D distributions of the surface density obtained via equation (\ref{eq:Sigma}) with some of the $\Sigma_d$ and $e_d$ profiles listed above. One can see that $\Sigma(r,\phi)$ can have a rather complicated structure, depending on the particular disk model used. The contours of constant $\Sigma(r,\phi)$ (on the left) are very different from the elliptical trajectories (on the right) of the mass elements giving rise to the surface density distribution in the disk.


\section{Results}  
\label{sect:results}

\begin{figure}
\centering
\includegraphics[width=0.5\textwidth]{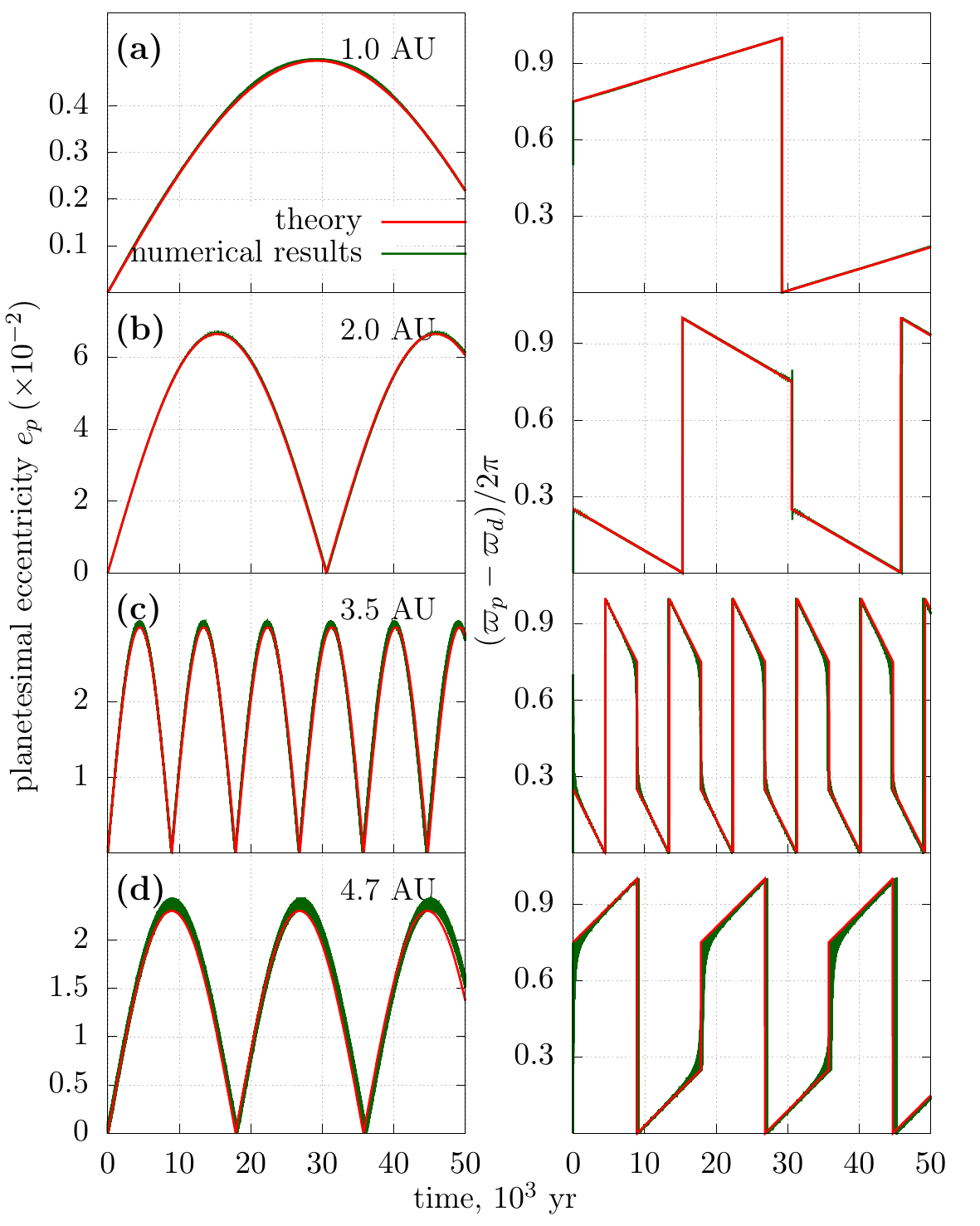}
\caption{
Verification of our analytical calculation of the disk disturbing function $R_d$ using direct numerical integration of particle orbits in the (numerically computed) disk potential with MERCURY. The disk model has $\Sigma_d(a)$ given by (\ref{eq:QuadSigma}) with $\Sigma_0\approx 1100$ g cm$^{-2}$ and $e_d(a)$ given by (\ref{eq:LinEcc}) with  $e_0=0.01$. The time evolution of the particle eccentricity $e_p$ (left) and apsidal angle $\varpi_p$ (right) is shown for different values of the particle semi-major axis $a_p$, as labeled on the panels. In all cases, test particles start with zero eccentricity, which results in pronounced secular oscillations of $e_p$, well described by the solution (\ref{eq:ecc_circ}). The tree eccentricity vector precesses at a steady rate, resulting in $\varpi_d$ evolution characterized by the solution (\ref{eq:ecc_circ}). 
\label{ex_orb}}
\end{figure}

We start our comparison by providing an illustration of the orbital evolution caused by the gravity of an underlying eccentric disk. Figure \ref{ex_orb} displays the variation of the test particle eccentricity $e_p$ (on the left) and apsidal angle $\varpi_p$ (on the right) in time, computed at several values of the semimajor axis $a_p$ for a particular disk model --- $\Sigma_d(a)$ given by equation (\ref{eq:QuadSigma}) with $\Sigma_0\approx 1100$ g cm$^{-2}$ and $e_d(a)$ given by equation (\ref{eq:LinEcc}) with $e_0=10^{-2}$. One can see that the agreement between the direct integration (green) and our secular prediction (red) is very good at all radii. Both $e_p(t)$ and $\varpi_p(t)$ follow the predictions of (\ref{eq:ecc_circ}) very closely, agreeing both in the amplitude of the eccentricity oscillations and in their phase (or period). 

From the behavior of $\varpi_p(t)$ at different semimajor axes, one can immediately see that  free precession of the particle orbit can be  {\it prograde} in some parts of the disk and {\it retrograde} in others. The change of sign of the precession occurs for orbits fully enclosed within the disk. It is also clear that not only the sign but also the period of the precession is a function of the location in the disk.

Evolutionary time series of $e_p(t)$, such as the one presented in Figure \ref{ex_orb}, allow one to easily measure the maximum amplitude of the eccentricity oscillations, $e^{\rm m}_p$, which should be compared to the theoretical values of $2B_d/A_d$; see equation (\ref{eq:ecc_circ}). Similarly, the period $P_{\rm sec}$ of each $e_p$ oscillation yields the corresponding free precession rate as $\dot \varpi_{\rm sec}=2\pi/P_{\rm sec}$, which in this Figure agrees very well with the theoretical $A_d$. The value of $\dot \varpi_{\rm sec}$ can also be independently inferred from the slope of the numerical $\dot\varpi(t)$ curves, which should be equal to $A_d/2$, see equation (\ref{eq:ecc_circ}).

Close inspection of Figure \ref{ex_orb} reveals additional features in the $e_p(t)$, $\varpi_p(t)$ evolution curves beyond the large-scale secular oscillations, which are well described by the solution (\ref{eq:ecc_circ}). These features manifest themselves as small-amplitude, short-period oscillations around the purely secular solution, most pronounced near the outer edge of the disk. The period of these oscillations is equal to the local orbital period $2\pi/n_p$. Their amplitude scales linearly with the disk mass, depends on the semimajor axis of test particle $a_p$,  but is independent of the disk eccentricity (we observe oscillations with the same amplitude even when the disk eccentricity is zero). We interpret these oscillations as resulting from the nonelliptical shape of the particle orbits in the combined potential of the star and the disk. As the numerical integration outputs osculating orbital elements (effectively fitting a pure ellipse at every point of a truly nonelliptical trajectory), oscillations of $e_p$ (as well as $\varpi_p$ and $a_p$) on a local dynamical timescale naturally arise. We did verify that the angular momentum of a test particle is strictly conserved through these oscillation cycles when $e_d=0$. A similar effect was discussed by \citet{Geo} in application to hierarchical triples.

In our subsequent presentation, we will focus on the behavior of $e^{\rm m}_p$ and $\dot \varpi_{\rm sec}$, derived from the data similar to that shown in Figure \ref{ex_orb}, in different disk models.

\subsection{Radial variation for different $\Sigma_d$ models}  
\label{sect:Sigma}


We now test the accuracy of our secular theory in disks with different $\Sigma_d(a)$ profiles. This allows us to examine the different possible behaviors of the secular coefficients $A_d$ and $B_d$ as well as to see how sensitive the agreement with the numerical results is to the different features of the $\Sigma_d(a)$ distributions.


\subsubsection{$\Sigma_d$ smoothly vanishing at the edges}  
\label{sect:quad}

\begin{figure}
\centering
\includegraphics[width=0.5\textwidth]{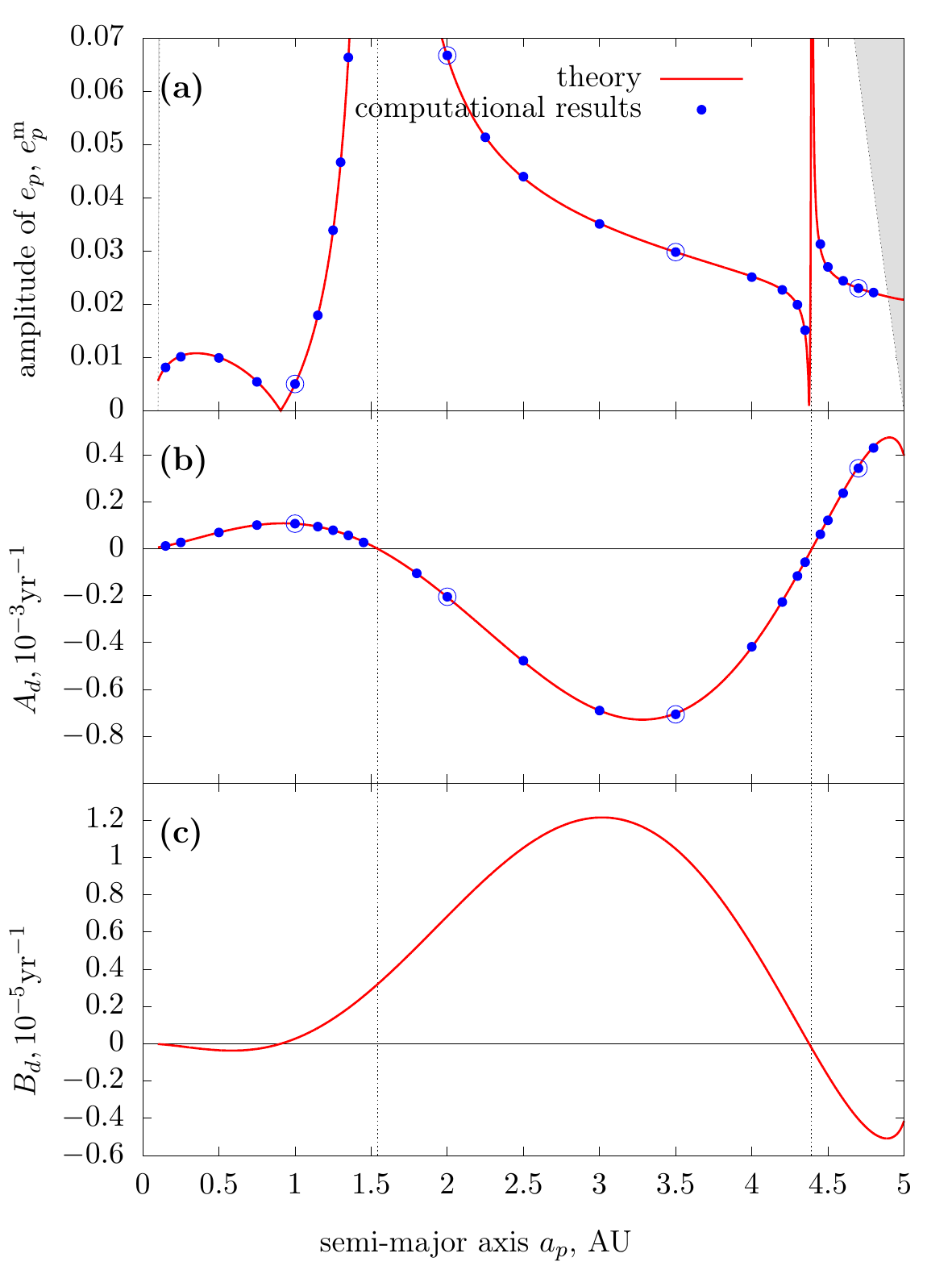}
\caption{
Characterization of secular oscillations of test particles (with orbits fully enclosed within the disk) driven by the disk gravity. Shown as a function of the semimajor axis of test particle $a_p$ are: (a) amplitude $e^{\rm m}_p$ of the eccentricity oscillations (blue dots) compared with theoretical $2e_{\rm forced} = 2B_d/A_d$ (red curve), (b) the frequency of secular oscillations $\dot \varpi_{\rm sec}$ (blue dots) compared against $A_d$ (red curve) and (c) the coefficient $B_d$ of the theoretical secular disturbing function. This calculation uses the disk model with $\Sigma_d$ distribution (\ref{eq:QuadSigma}) and $\Sigma_0\approx 1100$ g cm$^{-2}$, so that $\Sigma_d$ goes to zero at finite semimajor axes (0.1 AU and 5 AU), but smoothly, so that the boundary terms in the expressions for $A_d$ and $B_d$ do not arise; see \S \ref{sect:AdBd}. The radial profile of the disk eccentricity $e_d(a)$ is given by equation (\ref{eq:LinEcc}) with $e_0=0.01$. Circled dots correspond to the four values of $a_p$ used in Figure \ref{ex_orb}. Vertical dotted lines mark the locations where $A_d=0$. Gray regions near the disk edges correspond to the edge noncrossing constraints $a_p(1+e^{\rm m}_p)<a_2$ at apastron and $a_p(1-e^{\rm m}_p)>a_1$ at periastron. One can see excellent agreement between the secular theory and the direct numerical integration.
\label{fig:AB_SQ_radial}}
\end{figure}

We start by looking at the secular effect of a disk, in which $\Sigma_d$ smoothly goes to zero at the boundaries, i.e. the one with $\Sigma_d$ given by the equation (\ref{eq:QuadSigma}). In Figure \ref{fig:AB_SQ_radial} we display the theoretical radial profiles of $A_d$, $B_d$, and $2B_d/A_d$ for such a disk. The curves of $A_d(a_p)$ are compared with the numerically determined $\dot \varpi_{\rm sec}$ (blue dots), while the theoretical $2B_d(a_p)/A_d(a_p)$ is compared against the numerical $e^{\rm m}_p$. This particular calculation uses the linear eccentricity profile $e_d(a)$ given by the equation (\ref{eq:LinEcc}) with relatively low $e_0=0.01$, so that the maximum value of the disk eccentricity (reached at its inner edge) is 0.02. Thus, the requirement $e_d\ll 1$ necessary for the secular results (\ref{eq:R_gen}), (\ref{eq:AdBd})-(\ref{eq:Bd_edge}) to apply is fulfilled.

One can see an almost perfect match between the numerical results and theory, both in terms of the amplitude of the disk-induced particle eccentricity as well as the period of the associated orbital precession. Theoretical calculation easily reproduces even the very subtle features of the $e^{\rm m}_p(a)$ behavior, including the variations happening on very short radial scales manifesting themselves as sharp features in Figure \ref{fig:AB_SQ_radial}. There are several features to note in this figure.

First, $A_d$ has a different sign in different parts of the disk: precession of the free eccentricity vector is {\it prograde} near the boundaries of the disk, while it is {\it retrograte} away from the edges. This change of sign of $\dot \varpi_{\rm sec}$ was clear already in Figure \ref{ex_orb} (which is drawn for the same disk model as Figure \ref{fig:AB_SQ_radial}, at the locations highlighted with circles in the latter), but the Figure \ref{fig:AB_SQ_radial} provides a much more detailed representation of the different characteristics of the secular effect of the disk.

Second, $B_d$ also changes sign as $a_p$ varies. As a result of sign variations of both $A_d$ and $B_d$, the forced eccentricity vector ${\bf e}_{\rm forced}$ can be {\it aligned} with the apsidal line of the disk in some intervals of $a_p$, and {\it anti-aligned} with it in others. 

Third, both numerical and analytical $e^{\rm m}_p$ exhibit formal singularity at two distinct locations in this disk, $a\approx 1.54$ AU and $\approx 4.39$ AU. The origin of these singularities can be traced to the expression (\ref{eq:e_forced}) for the forced eccentricity (recall that our theory predicts $e^{\rm m}_p=2e_{\rm forced}$), which has $A_d$ in the denominator, and the fact that $A_d$ crosses zero (as the sense of ${\bf e}_{\rm free}$ precession changes) at these locations, see Figure \ref{fig:AB_SQ_radial}b. We will discuss these singularities in more detail in \S \ref{sect:singularity}.

Fourth, at certain values of the semimajor axis (at $\approx 0.92$ AU and $\approx 4.39$ AU) the disk-induced forced eccentricity vanishes. This happens because $B_d$ changes sign at these locations so that $B_d\to 0$; see Figure \ref{fig:AB_SQ_radial}c and equation (\ref{eq:e_forced}). Interestingly, one of the semimajor axes where $e^{\rm m}_p\to 0$ ($a_p\approx 4.39$ AU) is located in the immediate proximity of the singularity of $e^{\rm m}_p$. This occurs because in this part of the disk, $A_d$ and $B_d$ go through zero at almost the same (but still slightly different) values of $a_p$. 

\begin{figure}
\centering
\includegraphics[width=0.5\textwidth]{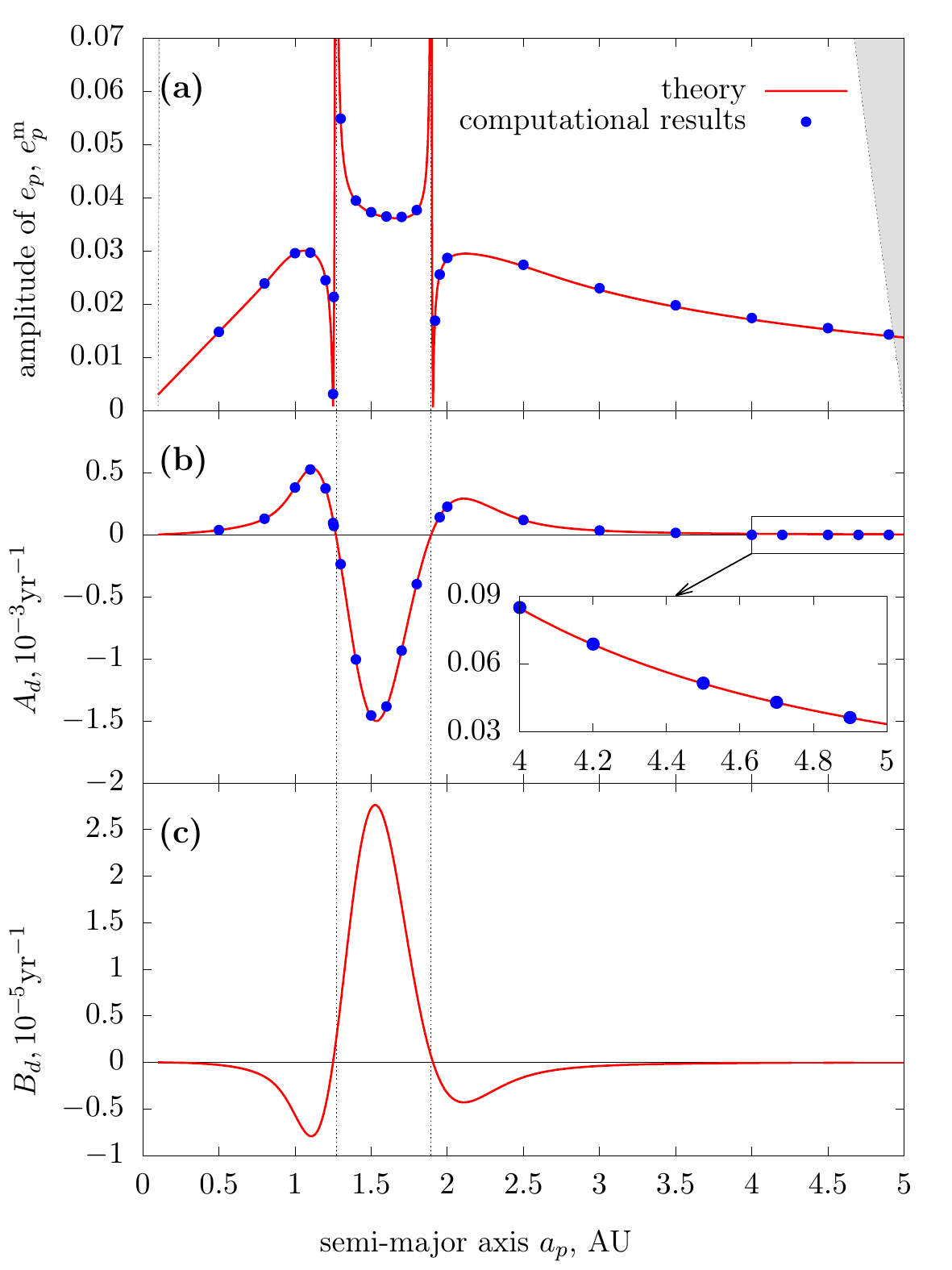}
\caption{
Same as Figure \ref{fig:AB_SQ_radial}, but now for a disk (Gaussian ring) model with different $\Sigma_d(a)$ distribution (\ref{eq:GaussSigma}), with $\Sigma_0=100$ g cm$^{-2}$; this model does not have boundaries at finite radii. The disk eccentricity profile is still given by equation (\ref{eq:LinEcc}) with $e_0=0.01$. The inset in panel (b) shows the behavior of $A_d$ far from the main body of the ring.
\label{fig:AB_Gauss_radial}}
\end{figure}

Fifth, for the surface density profile (\ref{eq:QuadSigma}), we find that $A_d$ and $B_d$ remain finite everywhere in the disk, including the boundaries. This is in line with the expectations outlined in \S \ref{sect:AdBd} for the disks with $\Sigma_d$ smoothly going to zero at the edges, which do not result in divergent boundary terms in $A_d$ and $B_d$.


\subsubsection{Disk without boundaries}  
\label{sect:Gauss}

Next we consider secular dynamics in the potential of a Gaussian ring with $\Sigma_d$  given by the equation (\ref{eq:GaussSigma}), which does not feature well-defined edges. We plot the behavior of the corresponding $e^{\rm m}_p$, $A_d$, and $B_d$ in Figure \ref{fig:AB_Gauss_radial}. 

One can see that many features of secular dynamics present in the case studied in \S \ref{sect:quad} are present here as well: both $A_d$ and $B_d$ are finite, they vary with $a_p$ and change sign, while $e^{\rm m}_p$ exhibits both singularities and nulls. This is likely related to the fact that these two $\Sigma_d$ profiles are morphologically similar: they exhibit no discontinuities, are single-peaked, and smoothly decay to zero away from the peak. The only obvious difference is that in the Gaussian case in Figure \ref{fig:AB_Gauss_radial} the nulls of $e^{\rm m}_p$ are located very close to {\it both} singularities of $e^{\rm m}_p$. However, the significance of this difference is unclear.


\subsubsection{$\Sigma_d$ sharply truncated at the edges}
\label{sect:sharp}

\begin{figure}
\centering
\includegraphics[width=0.5\textwidth]{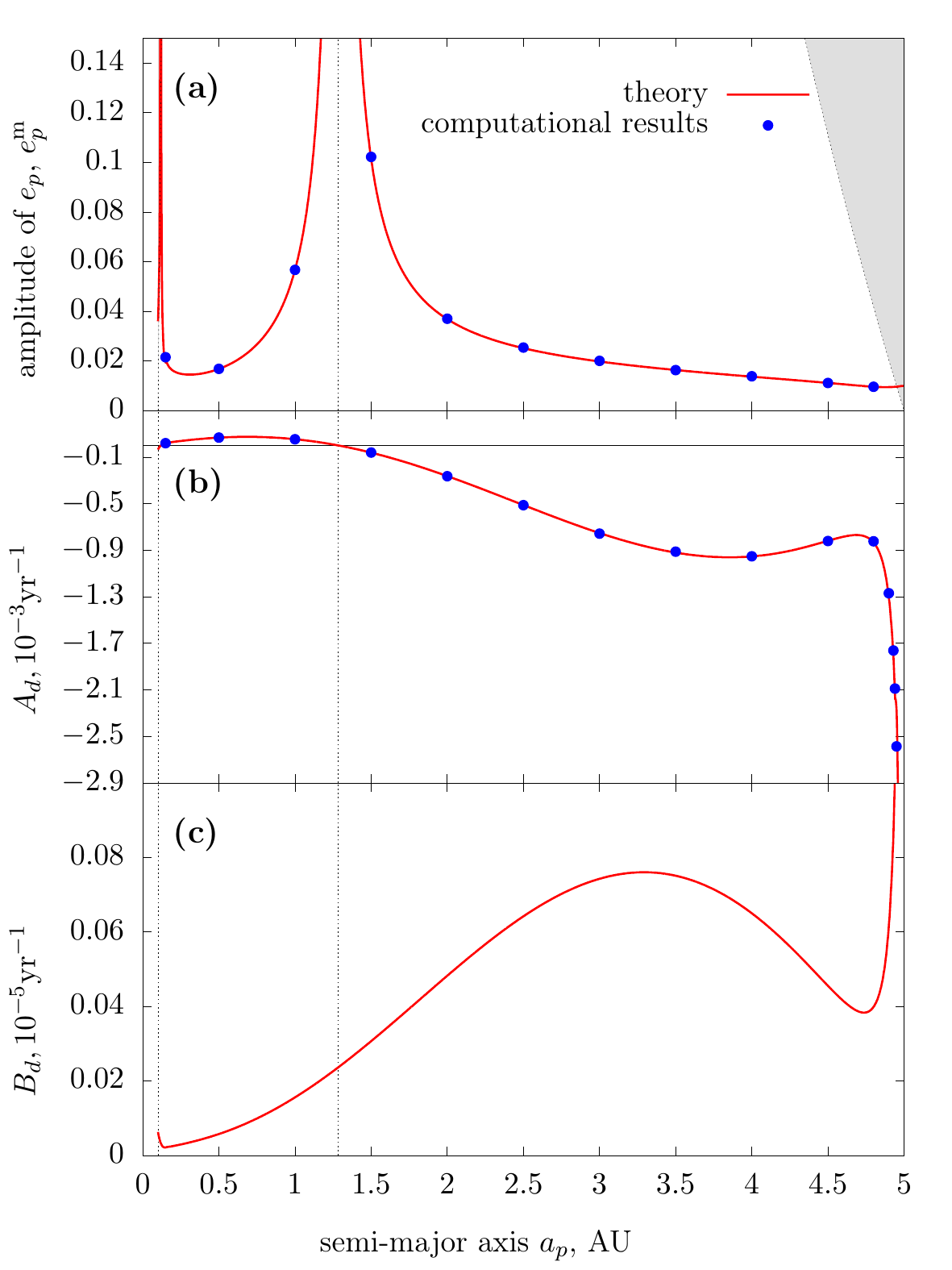}
\caption{
Same as Figure \ref{fig:AB_SQ_radial}, but now for the disk model with $\Sigma_d(a)$ distribution (\ref{eq:SinSigma}) and $\Sigma_0=20$ g cm$^{-2}$, which has sharp edges at finite semimajor axes (0.1 AU and 5 AU). The radial profile of the disk eccentricity is still given by equation (\ref{eq:LinEcc}), but now with the overall normalization $e_0=0.005$. Note the sharp increase of the amplitude of $A_d$ near the outer edge of the disk, accurately matched by our theory.
\label{fig:AB_Sin_radial}}
\end{figure}

We also examined secular behavior for the disk model in which surface density displays a discontinuous drop to zero at the inner and outer edges, namely, the one represented by equation (\ref{eq:SinSigma}). Sharply truncated $\Sigma$ distributions are rather typical in planetary rings and other astrophysical disks, and it is important to study the effect of their gravity on the dynamics of embedded objects. 

Figure \ref{fig:AB_Sin_radial} illustrates secular dynamics in the potential of such a disk. It again shows the variation of sign of $A_d$, which results in the emergence of two singularities of $e^{\rm m}_p$ --- one very close to the inner edge of the disk at $a_p=0.11$ AU and another at $a_p\approx 1.28$ AU. However, for this disk model, $B_d$ does not change sign --- it always stays positive. As a result, there are no nulls of ${\bf e}^{\rm m}_p$ for orbits in the potential of such a disk.

An even more dramatic difference with the cases explored in \S \ref{sect:quad}-\ref{sect:Gauss} is the behavior of $A_d$ and $B_d$ near the disk edges. Unlike the two previous cases, in which both coefficients remained finite everywhere at all radii, in a sharply truncated disk, $A_d$ and $B_d$ exhibit singularity as the disk edge is approached. This is very clearly seen at the outer\footnote{Numerical issues prevent us from demonstrating the divergent behavior of $A_d$ and $B_d$ at the inner edge of the disk.} boundary of the disk in Figure \ref{fig:AB_Sin_radial}, where both theoretical curves and the results of orbit integrations exhibit divergent behavior. Unfortunately, we cannot probe this divergence in great detail numerically, as the orbits of test particles start crossing the edge of the disk.  

At the same time, the test particle eccentricity $e^{\rm m}_p$ remains finite at the disk edges, even though both $A_d$ and $B_d$ are singular there. This is because both coefficients of the disturbing function (\ref{eq:R_gen}) diverge in similar fashion at the edge, so that $e_{\rm forced}$ remains finite there.

\begin{figure}
\centering
\includegraphics[width=0.5\textwidth]{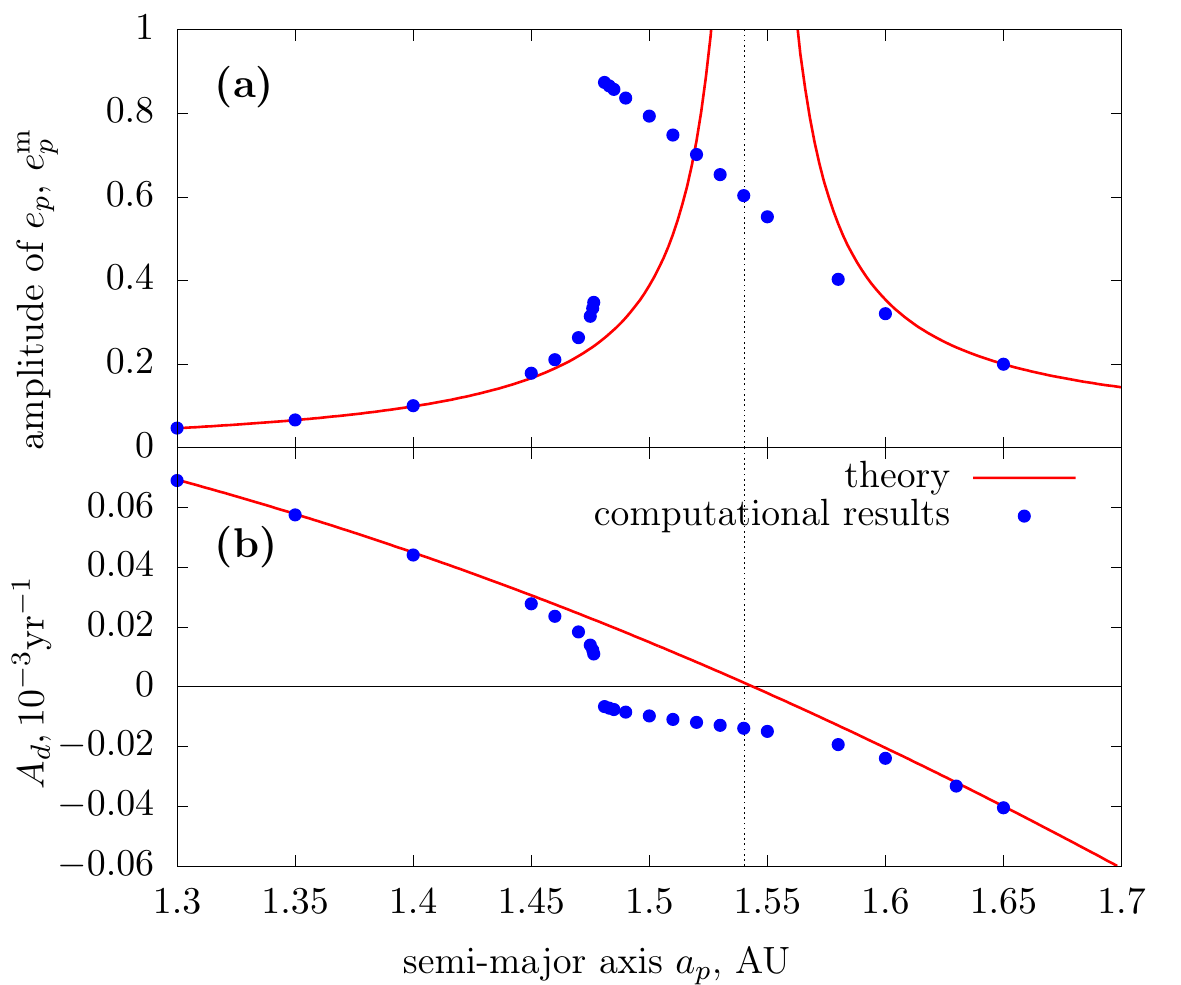}
\caption{
Zoom-in on a part of Figure \ref{fig:AB_SQ_radial} in the vicinity of a secular singularity at $a\approx 1.5$ AU. One can see how particle motion starts to deviate from our lowest-order secular theory as $e_p$ grows to values of order unity.
\label{fig:zoom}}
\end{figure}


\subsection{Singularities of $e_p^{\rm m}$}
\label{sect:singularity}


A common feature for all $\Sigma_d$ distributions examined in \S \ref{sect:Sigma} is the emergence of multiple singularities of $e^{\rm m}_p$. At these locations, $e^{\rm m}_p$ formally diverges, and the assumption $e_p\ll 1$ used in deriving our secular disturbing function (\ref{eq:R_gen})-(\ref{eq:R1}) breaks down. The way in which this happens is illustrated in Figure \ref{fig:zoom}, where we compare the numerical and analytical results in the vicinity of one of the singularities (near $a_p=1.54$ AU) in a disk with $\Sigma_d(a)$ given by equation (\ref{eq:QuadSigma}); see Figure \ref{fig:AB_SQ_radial}.

One can see that as the theoretical singularity is approached, the behavior of $\dot \varpi_{\rm sec}$ starts to deviate from the prediction (\ref{eq:AdBd})-(\ref{eq:Ad_edge}). As a result, $\dot \varpi_{\rm sec}$ goes through zero at a location slightly different from the one where $A_d=0$. Note that even at the point where $\dot \varpi_{\rm sec}=0$ particle eccentricity remains finite (even though it reaches values close to 1). This means that equation (\ref{eq:e_forced}) is no longer valid when $e_p\sim 1$ and that additional, higher-order terms become important in addition to the lowest-order secular potential contribution (\ref{eq:R_gen}). This discrepancy could be at least partly ameliorated by including higher-order (in $e_p$) terms in the calculation of $R_d$, as was done recently in \citet{Sefilian} for the case of power-law disks.

As evidenced by Figures \ref{fig:AB_SQ_radial} and \ref{fig:AB_Gauss_radial}, $e^{\rm m}_p$ singularities often occur in the immediate vicinity of nulls of $B_d$. This results in a characteristic shape of these singularities, with  $e^{\rm m}_p$ sharply dropping to zero in close proximity to the singularity. This leads to a dramatic difference in the eccentricities of particles with almost identical semimajor axes, resulting in their orbits crossing. Such locations thus provide a natural environment for particle collisions. 

\begin{figure}
\centering
\includegraphics[width=0.5\textwidth]{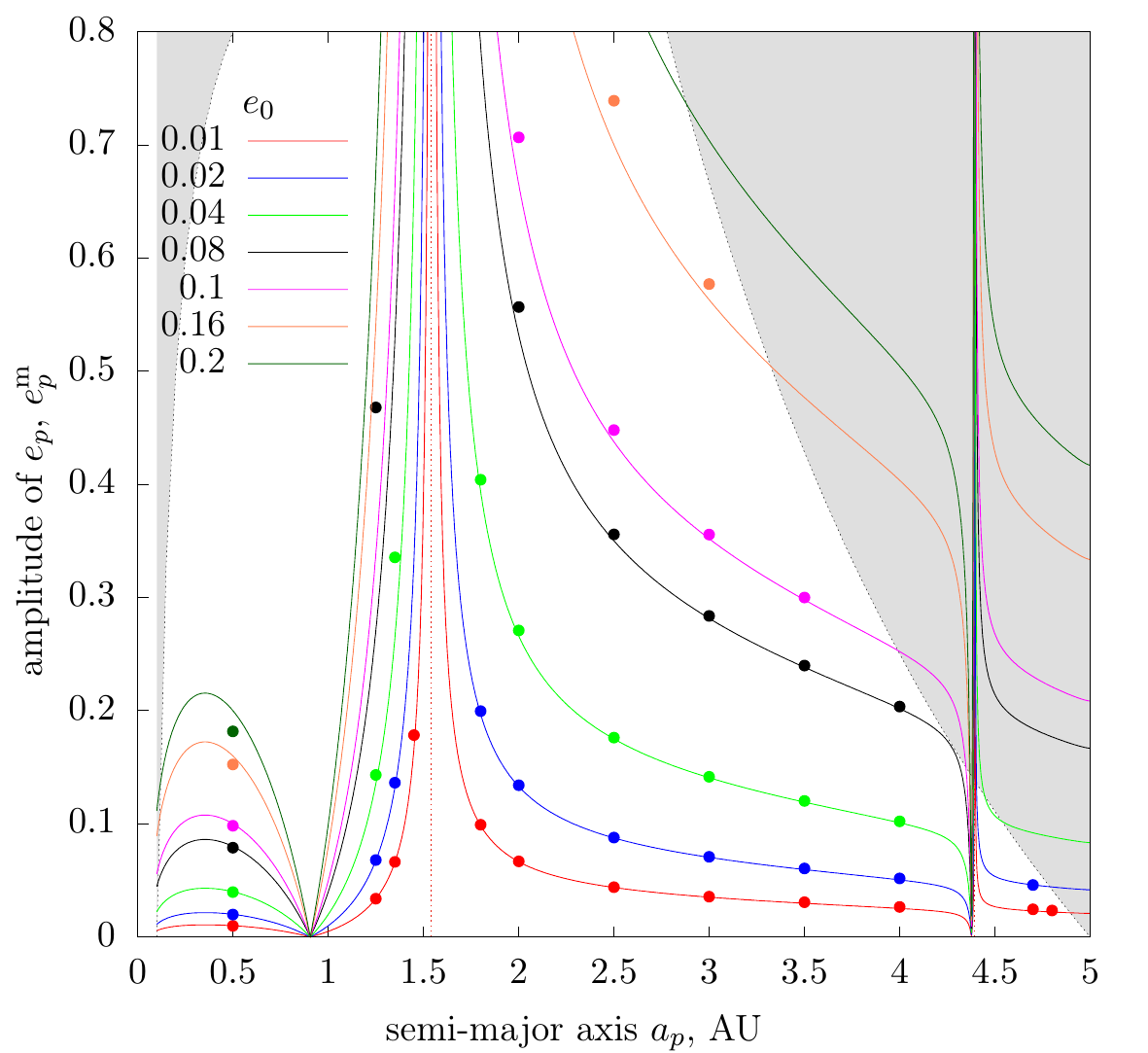}
\caption{
Agreement between our secular theory and direct orbit integrations as a function of the disk eccentricity amplitude. Shown is the amplitude of secular oscillations found in direct orbit integrations, as well as theoretical values of $2e_p^{\rm forced}$, shown as a function of both the distance from the star (horizontal axis) and the overall normalization $e_0$ of the disk eccentricity $e_d$ (different curves). This calculation assumes surface density profile in the form (\ref{eq:QuadSigma}) with $\Sigma_0\approx 1100$ g cm$^{-2}$ and the eccentricity profile given by equation (\ref{eq:LinEcc}). Continuous curves of different colors (corresponding to different values of eccentricity normalization $e_0$ in equation (\ref{eq:LinEcc}), as shown in the panel) represent analytical results based on our secular theory. Dots of different colors display the corresponding numerical results. Other notation is the same as in Figure \ref{fig:AB_SQ_radial}.
\label{fig:e_var_lin}}
\end{figure}

It is not clear why in some cases conditions $A_d=0$ and $B_d=0$ get realized at almost the same value of $a$. Inspection of the integrands in equations (\ref{eq:Ad_bulk}), (\ref{eq:Bd_bulk}) does not reveal an obvious reason for that to be the case. Interestingly, we find such "null-singularity" pairs only in the two disks without sharp edges (\S \ref{sect:quad}-\ref{sect:Gauss}), for which the boundary terms $A_d^{\rm edge}$ and $B_d^{\rm edge}$ in equations (\ref{eq:AdBd}) vanish. The disk with sharply truncated $\Sigma_d$ (see \S \ref{sect:sharp} and Figure \ref{fig:AB_Sin_radial}) has singularities without neighboring nulls of $B_d$ (in fact, $B_d$ does not change sign in this disk). Whether this outcome is due to the nontrivial boundary terms in this disk model is not clear. These curious properties of $e^{\rm m}_p$ singularities deserve further investigation.


\subsection{Sensitivity to the disk eccentricity $e_d$}
\label{sect:ecc}


Next we examine how our secular theory fares against changes of the disk eccentricity. First, we explore the effect of uniformly varying just the amplitude of eccentricity (\ref{sect:ecc_amp}), keeping the radial profile of $e_d(a)$ the same. We then look at the effect of a different radial profile of $e_d(a)$ on the agreement between the theory and numerical calculations (\S \ref{sect:ecc_prof}).


\subsubsection{Variation of the disk eccentricity amplitude}
\label{sect:ecc_amp}

In Figure \ref{fig:e_var_lin}, we plot $e^{\rm m}_p(a)$ for the disk model with $\Sigma_d$ and $e_d$ given by equations (\ref{eq:QuadSigma}) and (\ref{eq:LinEcc}), where we set $\Sigma_0\approx 1100$ g cm$^{-2}$ but vary the eccentricity normalization $e_0$ as indicated in the Figure (which is very similar to Figure \ref{fig:AB_SQ_radial}a). 

One can see that our theory works surprisingly well in predicting $e^{\rm m}_p(a)$ even when particle eccentricity reaches values in excess of $e_p=0.5$, when one would naively expect the description based on the lowest-order secular disturbing function (\ref{eq:R_gen}) to break down. For the $e_d$ profile (\ref{eq:LinEcc}), the maximum value of $e_d$, reached at the inner boundary of the disk, is $2e_0$. Thus, even for disks with the inner eccentricities reaching $e_d(a_1)=0.4$ our theory performs quite well.

\begin{figure}
\centering
\includegraphics[width=0.5\textwidth]{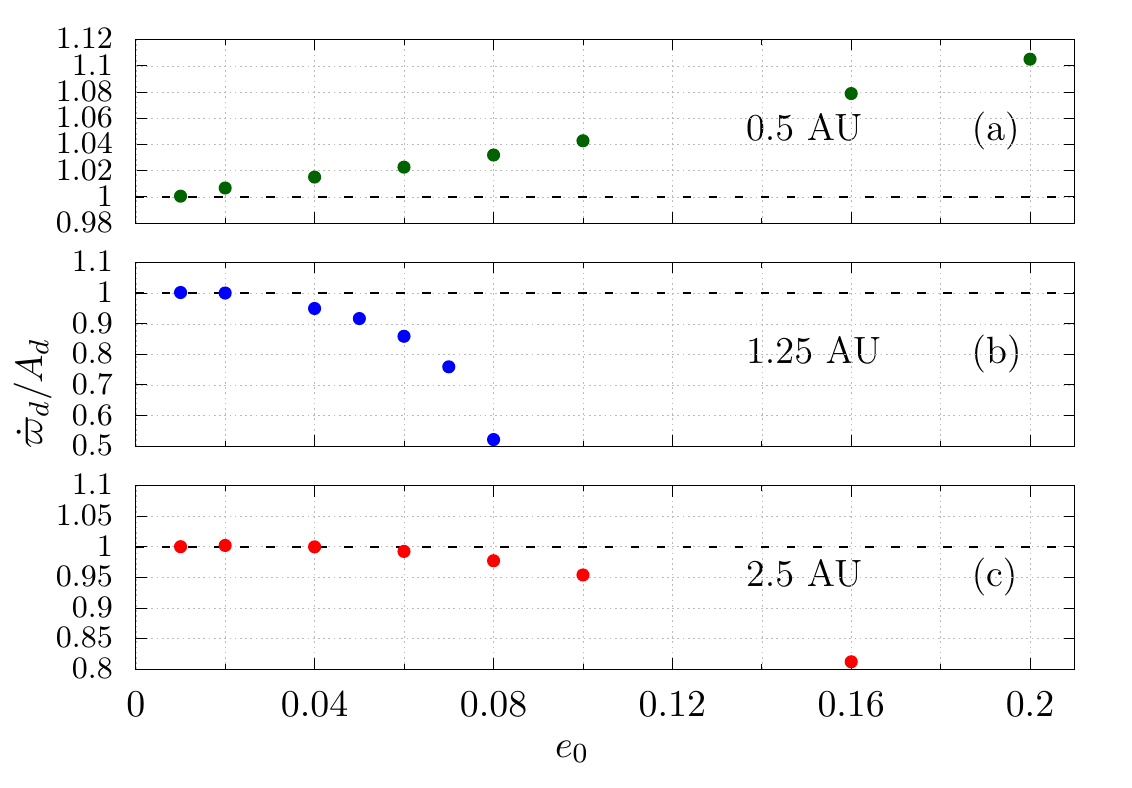}
\caption{
Relative deviation between the numerical ($\dot \varpi_{\rm sec}$) and theoretical ($A_d$) precession rates $\dot \varpi_{\rm sec}/A_d$, plotted as a function of the eccentricity amplitude $e_0$ of the linear disk eccentricity profile given by equation (\ref{eq:LinEcc}). Different panels correspond to different semimajor axes of the test particle: (a) $a_p=0.5$, (b) $1.25$, (c) $2.5$ AU. The calculation assumes $\Sigma_d$ model (\ref{eq:QuadSigma}) with $\Sigma_0\approx 1100$ g cm$^{-2}$.
\label{fig:A_diff}}
\end{figure}

The amplitude of eccentricity oscillations $e^{\rm m}_p(a)$ is just one metric by which the performance of our theory can be judged. Another obvious one is the free precession rate $\dot \varpi_{\rm sec}$. In Figure \ref{fig:A_diff} we plot the ratio of $\dot \varpi_{\rm sec}$ to its analytical counterpart $A_d$ as a function of the disk eccentricity normalization $e_0$. In the framework of our secular calculation, the precession rate should not depend on $e_0$ and be equal to $A_d$.

Figure \ref{fig:A_diff} shows that this is not really the case and $\dot \varpi_{\rm sec}$ does deviate from $A_d$ when $e_0$ is nonnegligible. The agreement between these two frequencies is somewhat less impressive than for $e^{\rm m}_p(a)$, with $\dot \varpi_{\rm sec}$ deviating from $A_d$ by tens of percent already for $e_0=0.1$. Note that the discrepancy between the numerical and analytical secular frequencies is a strong function of the semimajor axis $a_p$, with the largest deviation occurring in the vicinity of the $\dot \varpi_{\rm sec}$ singularity (just interior of it) at 1.3 AU. Thus, one should be careful when applying our lowest-order secular calculation to characterize the orbital evolution of test particles at certain locations in the disk with $e_d\gtrsim 0.1$.


\subsubsection{Variation of the disk eccentricity profile}  
\label{sect:ecc_prof}


\begin{figure}
\centering
\includegraphics[width=0.5\textwidth]{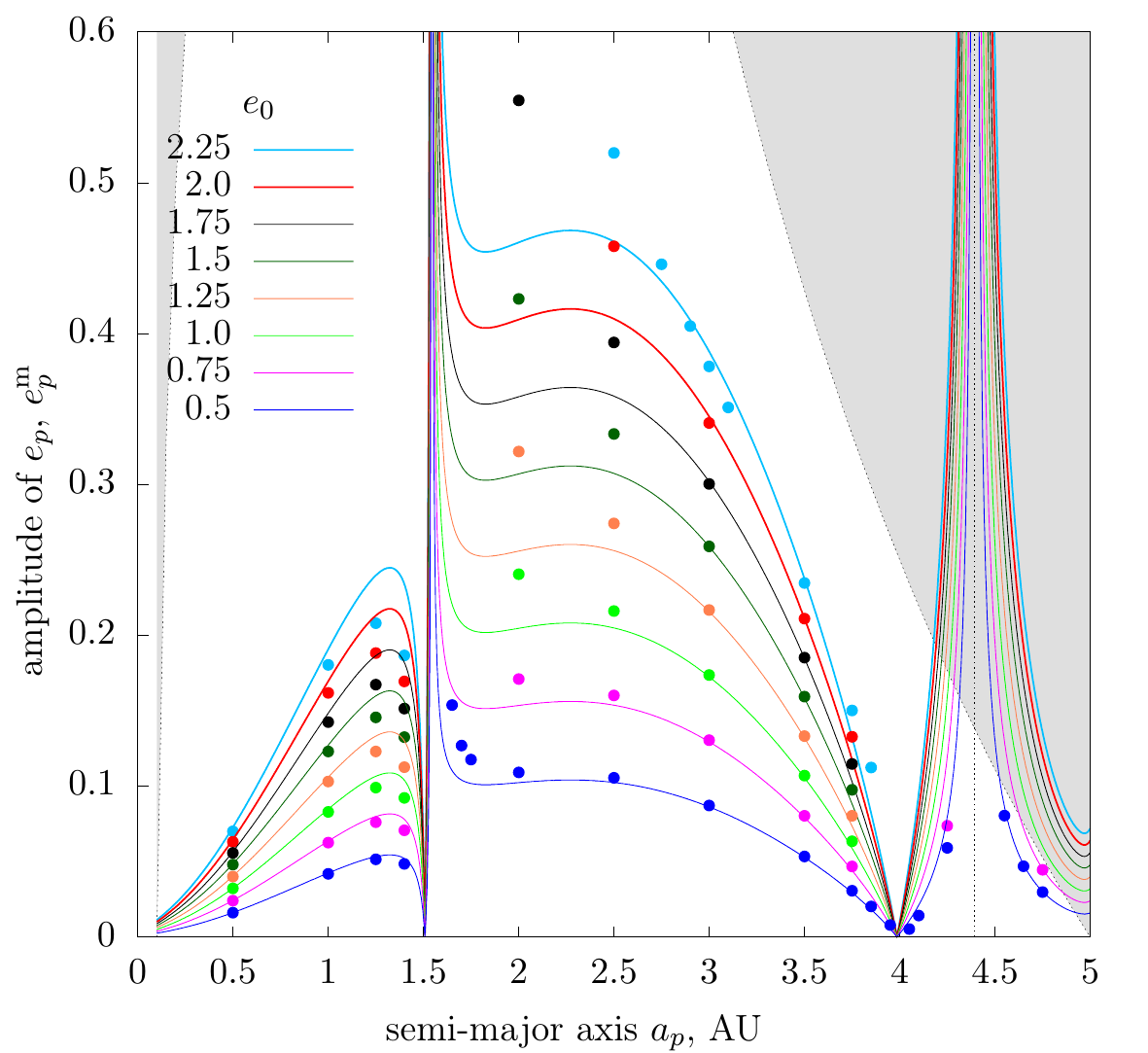}
\caption{
Same as Figure \ref{fig:e_var_lin} but now for a different eccentricity profile given by equation (\ref{eq:SqEcc}), with different $e_0$ corresponding to this profile. Comparing the results with Figure \ref{fig:e_var_lin} we see that variation of the  $e_d(a)$ profile changes the behavior of the maximum particle eccentricity, although the general topology of the curves remains roughly the same. The agreement between theory and direct orbit integrations is somewhat worse in this case compared to Figure \ref{fig:e_var_lin}.
\label{fig:e_var_quad}}
\end{figure}

Next we study the effect of changing the radial profile of the disk eccentricity $e_d(a)$. Figure \ref{fig:e_var_quad} is the analogue of Figure \ref{fig:e_var_lin} but made for a disk with an eccentricity profile (\ref{eq:SqEcc}). 

The first thing to note is that the radial profile of the theoretical $e^{\rm m}_p(a)$ still shows two clear singularities at the same radial locations as in Figure \ref{fig:e_var_lin}. This is easy to understand, since $e^{\rm m}_p$ diverges at radii where $A_d\to 0$, and $A_d$ is independent of the disk eccentricity profile (it depends only on the $\Sigma_d(a)$ profile). As a result, singularities of $e^{\rm m}_p$ stay at fixed locations even though $e_d(a)$ varies.  

Second, the detailed shape of $e^{\rm m}_p(a)$ is notably different from that shown in Figures \ref{fig:AB_SQ_radial} \& \ref{fig:e_var_lin}, despite the fact that the $\Sigma_d(a)$ profile is the same in both cases. While previously $e^{\rm m}_p$ was dropping to zero right next to the outer singularity at $a_p\approx 4.39$ AU, in Figure \ref{fig:e_var_quad} this happens near the inner singularity at $\approx 1.5$ AU. Another null of $e^{\rm m}_p$, more distant from the singularity, has also swapped its location and now lies in the outer part of the disk in Figure \ref{fig:e_var_quad}. 

Third, the agreement between the theory and direct orbit integrations is somewhat worse for the disk with an $e_d(a)$ profile (\ref{eq:SqEcc}). This is especially noticeable to the right of the inner singularity. For example, the $e^{\rm m}_p$ differs by $\approx 50\%$ from the theoretical expectation at $a_p=2$ AU for $e_0=1.75$ (black curve, which corresponds to the maximum eccentricity in the disk of $\approx 0.1$). Away from the region $1.5-2$ AU the agreement is generally better, even though the particle eccentricity excited by the disk potential can be quite high, $e_p\gtrsim 0.1-0.2$. 

We hypothesize that the reduced accuracy with which our secular theory predicts secular dynamics in the case of a disk with an $e_d(a)$ profile (\ref{eq:SqEcc}) is caused by the fact that this disk features a rather nonaxisymmetric surface density distribution $\Sigma(r,\phi)$. Indeed, Figure \ref{fig:green_maps}c,d show that the disk with $e_d(a)$ given by (\ref{eq:SqEcc}) features a well-defined concentration of mass in the outer region near the apastra of the constituent trajectories. As a result, for certain values of $a$ surface density $\Sigma(r_d,\phi_d)$ exhibits substantial variations {\it along the eccentric trajectories} of the mass elements comprising the disk, see Figure \ref{fig:green_maps}d. According to equation (\ref{eq:Sigma}), this is only possible if $\zeta e_d$ is not small for these values of $a$, meaning that the assumption $|\zeta e_d|\ll 1$ underlying our expansion (\ref{eq:Sig_expand}) is not fulfilled, which explains the deviations seen for this particular disk model.

\begin{figure}
\centering
\includegraphics[width=0.5\textwidth]{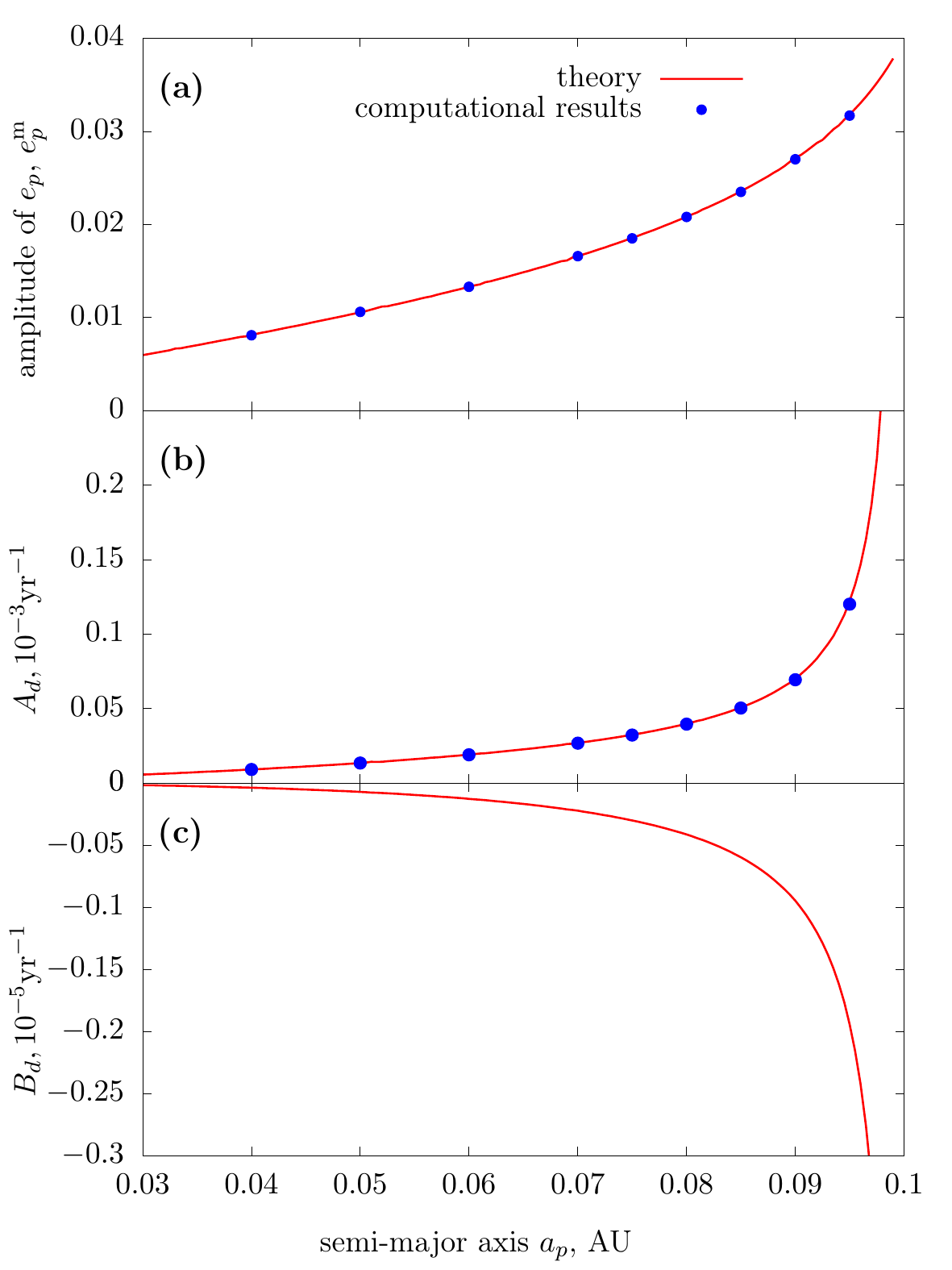}
\caption{
Characterization of the secular behavior in the potential of a disk with the same parameters as in Figure \ref{fig:AB_Sin_radial}, but now computed for a test particle orbiting {\it inside} the inner hole of the disk at 0.1 AU (i.e. outside the main body of the disk with sharp edges). One can see that our theory still works very well even beyond the radial extent of the disk. 
\label{fig:AB_SIN_inner}}
\end{figure}

\begin{figure}
\centering
\includegraphics[width=0.5\textwidth]{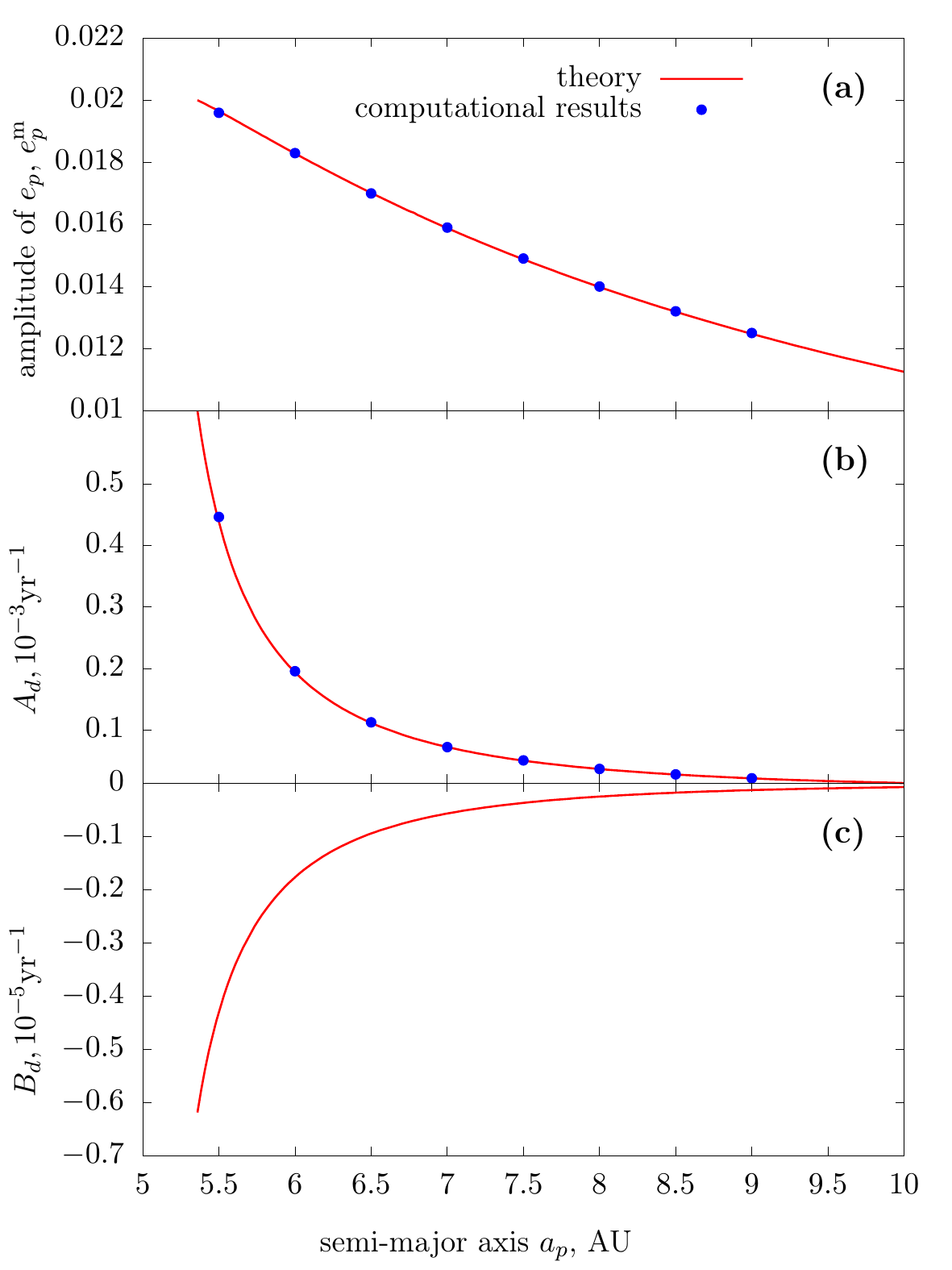}
\caption{
Same as Figure \ref{fig:AB_SIN_inner}, but now for a test particle orbiting {\it outside} the outer edge of the disk at 5 AU.
\label{fig:AB_SIN_outer}}
\end{figure}

By contrast, the disk with the eccentricity profile (\ref{eq:LinEcc}) shows a more axisymmetric distribution of $\Sigma(r,\phi)$, see Figure \ref{fig:green_maps}a,b, meaning smaller $|\zeta e_d|$ and higher accuracy of the expansion (\ref{eq:Sig_expand}). Thus, we expect that our secular theory should perform better for disks without highly nonuniform azimuthal features in $\Sigma(r_d,\phi_d)$.


\subsection{Motion outside the disk}  
\label{sect:outside}


Our derivation of the secular disturbing function in Appendix \ref{sect:disk_potential} explicitly assumes that the particle orbits fully {\it within} the disk. In other words, in the case of a disk with $\Sigma_d$ dropping to zero at some finite $a_1$ and $a_2$ the particle semimajor axis $a_p$ satisfies $a_1<a_p<a_2$, and its eccentric orbit does not cross the boundaries of the disk. Obviously, if the disk has no edges, as in, e.g., the case of a Gaussian ring (\ref{eq:GaussSigma}), then the particle orbit is always fully enclosed within the disk.

However, close examination of our derivation of $R_d$ in Appendix \ref{sect:disk_potential} demonstrates that it should also apply equally well {\it outside} the disk with sharp edges, as long as the particle orbit does not cross the disk boundaries. For example, in the case of a particle orbiting  outside the outer edge of the disk ($a_p>a_2$), first, one drops the contribution of the outer disk (i.e. $a>a_p$ as there is no disk material there) when computing $R_d$, and, second, the integration in the inner disk runs not up to $\alpha=1$ but only up to $\alpha=a_2/a_p<1$. As a result, the resultant expressions (\ref{eq:AdBd})-(\ref{eq:Bd_edge}) for the coefficients $A_d$ and $B_d$ apply without modification, even when $a_p<a_1<a_2$ or $a_1<a_2<a_p$.

To verify this claim, in Figures \ref{fig:AB_SIN_inner} \& \ref{fig:AB_SIN_outer}, we show radial profiles of $e^{\rm m}_p$ and $\dot \varpi_{\rm sec}$, as well as those of $2e_{\rm forced}$, $A_d$ and $B_d$ for particles orbiting inside and outside (correspondingly) the radial extent of the disk. Disk parameters are the same as in Figure \ref{fig:AB_Sin_radial}; in particular, $\Sigma_d$ is given by (\ref{eq:SinSigma}) with sharp edges at $a_1=0.1$ AU and $a_2=5$ AU. 

One can see that, as in the previous cases illustrated in Figures \ref{fig:AB_SQ_radial}-\ref{fig:AB_Sin_radial}, there is excellent agreement between our theory and direct orbit integrations, as long as the disk and particle eccentricities are low. Both the theoretical and the numerical values of the particle eccentricity extrapolated to the disk edges match the corresponding values extrapolated from inside the disk; see Figure \ref{fig:AB_Sin_radial}. At the same time, both $A_d$ and $B_d$ diverge as $a_p\to a_1-0$ and $a_p\to a_2+0$, mirroring the singularity of these coefficients identified previously (\S \ref{sect:AdBd}).

These results extend the applicability of our calculation of $R_d$ to (coplanar) particles having {\it arbitrary} semimajor axis relative to the disk, as long as their orbits do not cross the edge of the disk where $\Sigma_d$ discontinuously drops to zero. However, it can be shown that even this latter constraint can be removed, extending the applicability of our results even further. We do not dwell on this point here\footnote{Verification of this statement by direct orbit integrations can be tricky because of the formal logarithmic divergence of the acceleration at the edge of the disk, see \S \ref{sect:edges}.}, deferring it to a future study.


\section{Applications}  
\label{sect:apply}


The results presented in \S \ref{sect:results} demonstrate the validity and accuracy of the secular theory developed in this work in the low-$e$ limit. This motivates us to use this theory to further explore several aspects of secular motion in the potential of an eccentric disk. 

\begin{figure}
\centering
\includegraphics[width=0.5\textwidth]{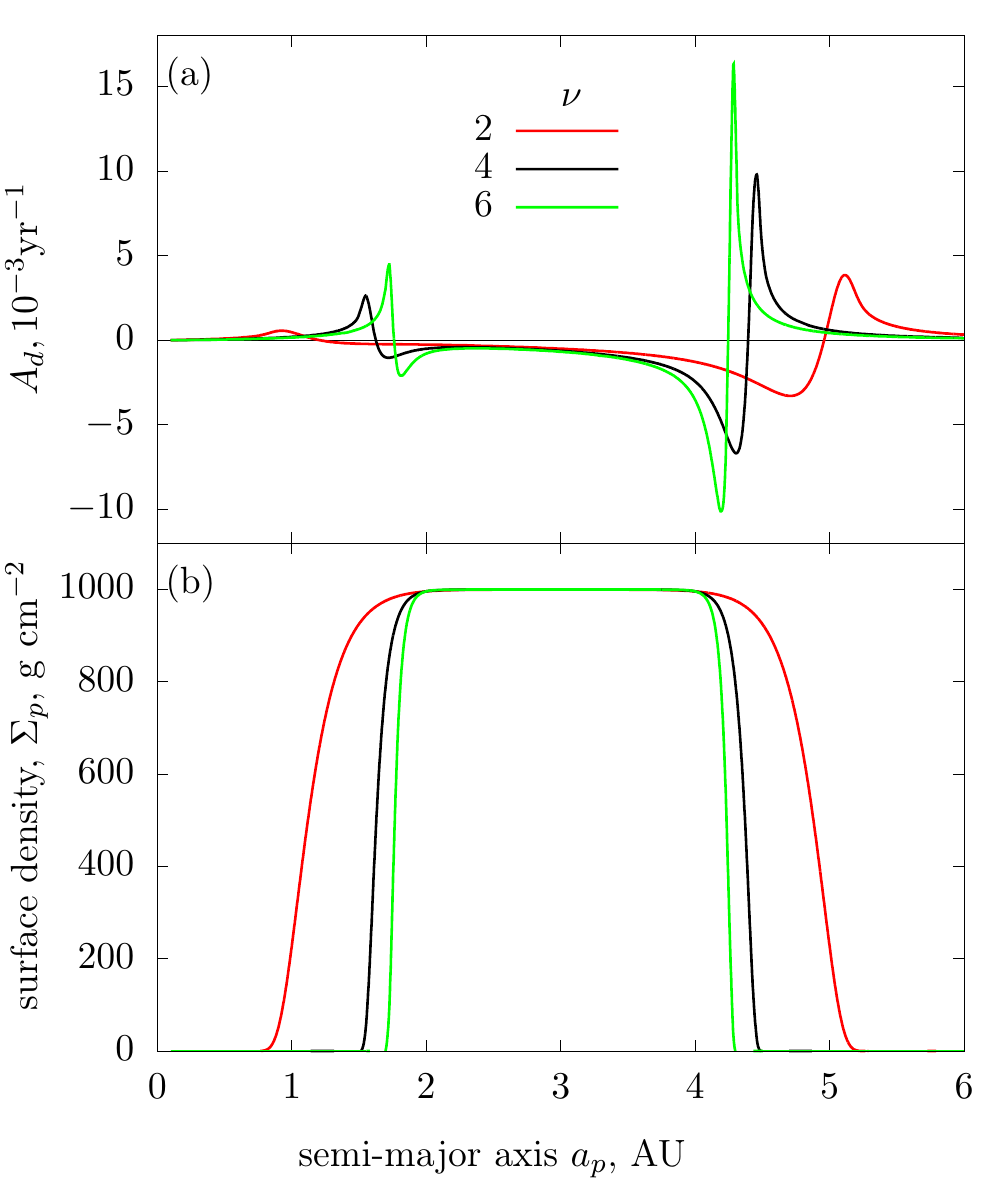}
\caption{
Illustration of the divergence of the free precession rate $A_d$ (panel (a)) near the sharp edges of the disk. A massive circular ring with three different profiles of $\Sigma_d$ given by equation (\ref{eq:ring}) and shown in panel (b), varying in sharpness of $\Sigma_d$ drop near the edges (regulated by the parameter $\nu$, as shown in panel (a)), is considered. One can see that the more abrupt the variation of $\Sigma_d$ is at the edges, the higher is the amplitude of $A_d$ that is reached in these regions. In the limit of a discontinuity in $\Sigma_d$ one would find $A_d\to \infty$, in agreement with our theory. 
\label{fig:edges}}
\end{figure}


\subsection{Edge effects}
\label{sect:edges}


Some astrophysical disks are known to have very sharp edges. For example, using {\it Voyager} 2 occultation data, \citet{Graps} demonstrated that the $\epsilon$ ring of Uranus has $\Sigma_d$ steeply dropping to zero at the ring boundaries. The edges of the Saturn rings are also known to be very sharp \citep{Tiscareno}. Outside the Solar System, eclipse data reveal the sharpness of the edge of the circumbinary ring around the young star KH 15D \citep{Winn}.   

Our calculations predict that sharp edges result in the divergence of the secular potential of a razor-thin disk, leading to a divergence in the precession rate near these locations (\S \ref{sect:sharp}). This outcome, resulting from the nonvanishing boundary terms, was previously pointed out in \citet{SR15b} for truncated power-law disks, and now we generalize it for other models of eccentric disks with edges. This prediction is nicely confirmed by the direct integration of particle orbits in a particular disk model with sharp edge; see Figure \ref{fig:AB_Sin_radial} and \S \ref{sect:sharp}. 

The divergence of $\dot \varpi_{\rm sec}$ (or $A_d$) near the sharp edge of a zero-thickness disk can be traced to the fact that the (in-plane) gravitational acceleration in this region behaves as $g_d\propto \ln|\Delta r|$, where $\Delta r$ is the separation from the edge. Specializing to the case of an axisymmetric disk, one then finds the free precession rate \citep{Fontana} 
\ba  
\dot \varpi_{\rm sec}=-\frac{n}{2 r a_{\rm K}}\frac{dg_d}{dr}\propto (\Delta r)^{-1}
\label{eq:precrate}
\ea  
(where $g_{\rm K}=GM_\star/r^2$) near the edge at the leading order. The divergent behavior $\dot \varpi_{\rm sec}\propto (\Delta r)^{-1}$ coincides with the scaling of the boundary terms in the expression (\S \ref{eq:Ad_edge}) for $A_d^{\rm edge}$, see (\ref{sect:AdBd}). Analogous  singularities should arise at any radius in the disk where $\Sigma_d(a)$ exhibits a discontinuity. 

A disk with $\Sigma_d$ dropping to zero smoothly over a narrow but finite range of $a$ would not have $A_d$ diverging there, as the boundary terms ($A_d^{\rm edge}$ and $B_d^{\rm edge}$) vanish for smooth $\Sigma_d$ profiles. Nevertheless, $A_d$ still exhibits a nontrivial behavior in this region. This is illustrated in Figure \ref{fig:edges} where we plot $\dot \varpi_{\rm sec}=A_d(a)$ (in the top) for rings with several $\Sigma_d(a)$ profiles\footnote{The explicit expression for $\Sigma_d(a)$ is given by the equation (\ref{eq:ring}).} of different degree of steepness near the boundaries (shown in the bottom). One can see that near the edge, $\dot \varpi_{\rm sec}$ exhibits very rapid variation, changing from large negative values just inside the edge to large positive values just outside the edge. 

This behavior can be understood by noticing that equation (\ref{eq:Ad_bulk}) for $A_d^{\rm bulk}=A_d$ contains a second derivative of the surface density $\Sigma_d$ inside the integral. Very close to the boundary, where $\Sigma_d$ is still high, $\Sigma_d^{\prime\prime}(a)$ is large and negative. Due to the logarithmic singularity of $b_{1/2}^{(0)}(\alpha)$ as $\alpha\to 1$, the radial convolution in equation (\ref{eq:Ad_bulk}) enhances the contribution of this region (with large $\Sigma_d^{\prime\prime}<0$) to $A_d$, resulting in {\it rapid retrograde} precession in this part of the disk. On the other hand, just slightly away from the boundary, where $\Sigma_d$ is low, $\Sigma_d^{\prime\prime}(a)$ is large and positive, driving {\it fast prograde} precession there.

As the sharpness of the $\Sigma_d$ profile near the boundaries increases, so does the magnitude of $\Sigma_d^{\prime\prime}(a)$. As a result, the amplitude of $\dot \varpi_{\rm sec}=A_d$ on both sides of the edge grows. In the limit of the infinitely sharp $\Sigma_d$ transition at the edge, the behavior of $A_d$ becomes singular, with the change of sign at the edge, in agreement with the expectation (\ref{eq:precrate}). In our formalism, this is accounted for mathematically by the appearance of the boundary terms (\ref{eq:Ad_edge}), whereas $\Sigma_d^{\prime\prime}$ in the integral (\ref{eq:Ad_bulk}) remains finite.   

In a disk with small but finite vertical thickness $h\ll a$ the behavior of the coefficients of $R_d$ would be slightly different. In such a disk, the rise of $A_d$ and $B_d$ would saturate at a finite value $\propto h^{-1}$ as the edge is approached. This transition is easy to understand by setting $\Delta r\sim h$ in equation (\ref{eq:precrate}).

The divergence of $A_d$ near the sharp edges can have important implications for, e.g., the dynamics of planetary rings. Even though $\Sigma_d$ remains finite at the edge, our results demonstrate that particle orbits should precess very rapidly and at a rapidly changing rate as their semimajor axes get closer to the edge. This naturally leads to particle orbit crossing resulting in their collisions, helping redistribute angular momentum near the disk edge \citep{CG2000,CC2003}.


\subsection{Free precession in disks with gaps}
\label{sect:gap}


\begin{figure}
\centering
\includegraphics[width=0.5\textwidth]{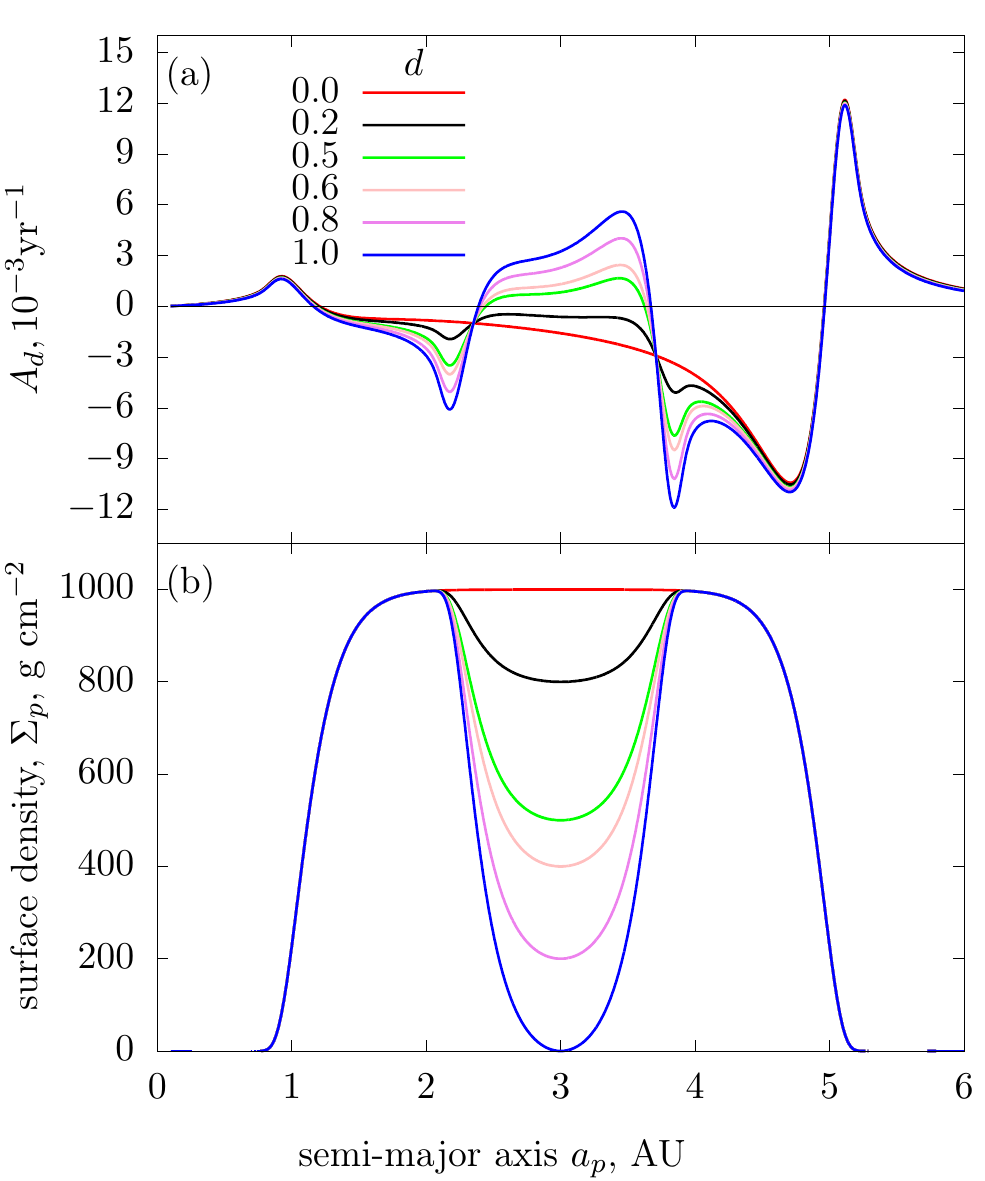}
\caption{
Variation  of the free eccentricity precession rate $A_d$ (panel (a)) in a disk with a gap. An underlying profile (\ref{eq:ring}) of $\Sigma_d$ is modified by imposing a gap of different relative depth $d$ (values are shown in panel (a)) according to the prescription (\ref{eq:Gap}), as illustrated in panel (b). The width of the gap w$=1.5$ AU is fixed. Note the evolution of the precession rate in the gap from negative to positive as the gap depth is increased.
\label{fig:gap_depth}}
\end{figure}

\begin{figure}
\centering
\includegraphics[width=0.5\textwidth]{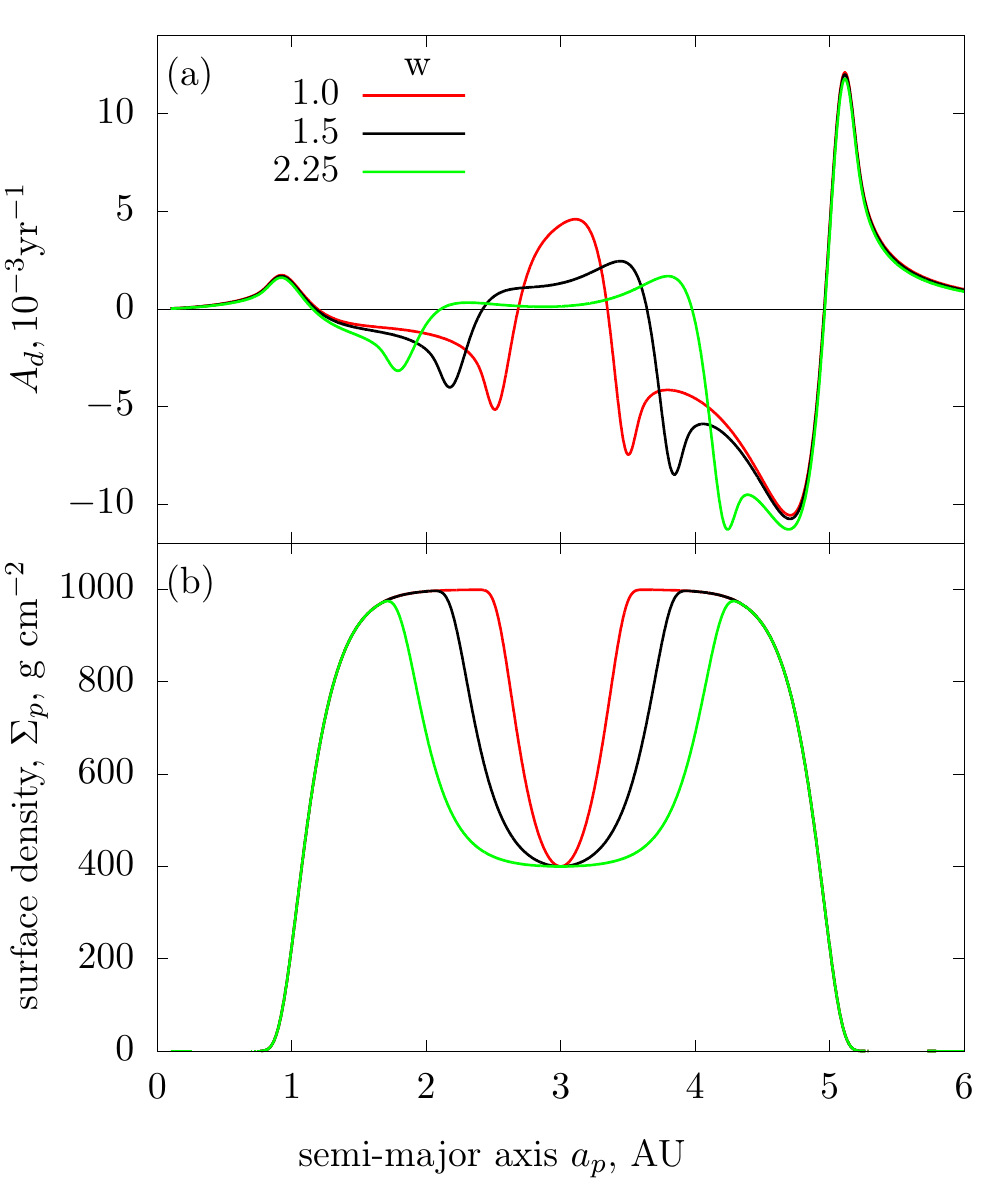}
\caption{
Same as Figure \ref{fig:gap_depth} but now showing the effect of the gap width w (indicated in panel (a)) on the behavior of $A_d$. The gap depth in equation (\ref{eq:Gap}) is kept fixed at $d=0.6$. 
\label{fig:gap_width}}
\end{figure}

Disk gaps present another example of sharp surface density gradients. Gaps could form naturally as a result of gas clearing by the gravitational torques due to massive planets orbiting within the disk. 

\citet{Ward} has looked at the effect of gaps on the free precession rate of test particles in axisymmetric disks with a power-law profile of $\Sigma_d$, finding $\dot \varpi_{\rm sec}$ to be negative within the disk, but positive within the gap. \citet{Ward} modeled the gap by setting $\Sigma_d$ to zero within a range of semimajor axes, which makes his result quite natural. Indeed, a disk with such a gap can be viewed as a combination of two disjoint disks with sharp edges. An object orbiting within a gap is thus moving exterior to an inner truncated disk and interior to an outer truncated disk. Our results in \S \ref{sect:sharp}, \ref{sect:outside}, \ref{sect:edges} demonstrate that interaction with each of the disks drives {\it prograde} free precession of such an object, with the combined effect being simply a linear superposition (also prograde) of the two contributions.

Using our theoretical formalism, we can explore how the results of \citet{Ward} change for more realistic (i.e. less sharp) gap profiles. We consider a disk, which is a combination of a wide ring with flat $\Sigma_d$ given by equation (\ref{eq:ring}), and a gap of width $w$ and relative depth $d$ (changing from $d=0$ for no gap to $d=1$ for $\Sigma_d=0$ in the gap center), with the profile given by equation (\ref{eq:Gap}). In Figure \ref{fig:gap_depth}, we show how $A_d=\dot \varpi_{\rm sec}$ varies as we change the gap depth. One can see that $A_d$ inside the gap region, which is negative in the absence of a $\Sigma_d$ depression, gradually decreases in magnitude, crosses zero, and becomes large and positive as the gap depth is increased. At the edges of the gap, $A_d$ exhibits a nontrivial structure reminiscent of that seen in Figure \ref{fig:edges}. If our gap had a more abrupt drop of $\Sigma_d$ at its edges, we would have converged to the case explored by \citet{Ward}.

Figure \ref{fig:gap_width} looks at the effect of variation of the gap width, keeping its depth constant. One can see that narrower gaps have a more pronounced effect on free precession in the center of the gap. This is because the edges of such gaps are sharper and also closer to the orbit of a precessing object. In this case, the intuition developed in \S \ref{fig:edges} again suggests that $A_d$ should be large and positive, as is observed in Figure \ref{fig:gap_width}. To summarize, the effect of a gap on free precession seems to be determined more by the {\it sharpness} of the $\Sigma_d$ gradient rather than by the gap width or depth separately.


\subsection{Free precession in smooth disks}
\label{sect:prec}


The nontrivial behavior of $\dot \varpi_{\rm sec}$ discussed in \S \ref{sect:edges}-\ref{sect:gap} is caused by the localized sharp features in $\Sigma_d(a)$. However, our results in \S \ref{sect:quad}-\ref{sect:sharp}, confirmed by direct orbit integration, clearly demonstrate that $A_d$ can also easily change sign within a disk with a {\it smooth} distribution of $\Sigma_d$. Since, according to the solution (\ref{eq:efree}), $A_d=\dot \varpi_{\rm sec}$ is the rate of precession of the free eccentricity vector, this implies that free precession of a test particle can be {\it both prograde and retrograde}, depending on its location within the disk. 
The possibility of a change of sign of free precession may appear rather surprising, given that the simple truncated power-law models usually employed to study the gravitational effect of thin disks tend to predict $\dot \varpi_{\rm sec}<0$, i.e. {\it retrograde} free precession \citep{Heppen,Ward,Batygin2011}. Prograde precession is (locally) possible in disks featuring gaps, i.e. sharp drops of $\Sigma_d$ (\citealt{Ward}; see also \S \ref{sect:gap}), but the disks explored in \S \ref{sect:quad}-\ref{sect:Gauss} have rather smooth profiles of $\Sigma_d(a)$. 

On the other hand, already in \citet{SR15b} it was shown that even the truncated power-law disks can exhibit {\it prograde} free precession far from the disk edges for certain values of the density slope $p=-d\ln\Sigma_d/d\ln a$, e.g. when $p<0$ so that $\Sigma_d$ grows with $a$. Given that near the edges (but still within the disk) one expects $A_d<0$ (see \S \ref{sect:edges}), the direction of free precession must change sign at some location inside such a disk. In light of this observation, our current results simply show that the change of sign of free precession is a rather common phenomenon for arbitrary profiles of $\Sigma_d$. 

Unfortunately, predicting the direction of the free precession (i.e. the sign of $A_d$) at a specific semimajor axis $a$ in a disk with a given profile $\Sigma_d(a)$ is not an easy task. Even in a disk without sharp boundaries, when the boundary term $A_d^{\rm edge}$ vanishes and $A_d=A_d^{\rm bulk}$ is represented by the equation (\ref{eq:Ad_bulk}), it is generally not straightforward to predict a priori the sign of this integral term. Indeed, a smooth, continuous $\Sigma_d(a)$ gradually decaying to zero at finite (e.g. given by equation (\ref{eq:QuadSigma})) or infinite (e.g. profile (\ref{eq:GaussSigma})) boundaries would necessarily have $\Sigma_d^{\prime\prime}(a)$ changing sign within the disk. Integration over $a$ provides a nontrivial, nonlocal mapping between the global behavior of $\Sigma_d^{\prime\prime}(a)$ and the value (and sign) of $A_d$.

As discussed in \S \ref{sect:Sigma} and \ref{sect:singularity}, a change of sign of $A_d$ inside the disk is also important because it gives rise to very high (formally divergent) values of the test particle eccentricity at radii where $A_d=0$, as long as the disk eccentricity $e_d$ is nonzero. Previously, a similar effect --- a localized singularity of $e_p$ --- was identified in studies of planet formation within stellar binaries, both analytically \citep{Rafikov13a,Rafikov13b,RS15a,RS15b,SR15a,SR15b} and numerically \citep{Meschiari}. Its origin could be traced to a secular resonance, caused by the cancellation of prograde precession due to the binary companion and retrograde precession driven by the disk gravity in presence of the nonzero binary torque (and disk torque, if the disk is eccentric). The importance of these singularities for planet formation in binaries lies in the fact that high values of $e_p$ lead to very energetic collisions between planetesimals, resulting in their destruction and hampering planetary accretion. In this work, we show that the same mechanism works even without a binary companion --- disk torque is always present when $e_d\neq 0$ and results in divergent $e_p$ whenever $A_d\to 0$, which naturally happens in our disks. 

As shown in \citet{RS15b} and \citet{SR15a}, an opposite effect is also possible in disks in binaries --- at certain locations, the eccentricities of test particles affected by the combined potential of a companion and an eccentric disk become very small, as a result of the cancellation of the corresponding torques\footnote{This often requires a particular relative orientation of the disk and binary apsidal lines \citep{RS15b}.}. Again, in our case, this happens even without a binary companion --- at the locations where $B_d=0$, one naturally finds $e_p\to 0$, as a result of the cancellation of torques arising from different parts of the same disk. This can be seen, e.g., at $\approx 0.9$ AU in Figures \ref{fig:AB_SQ_radial}, \ref{fig:e_var_lin} and at $\approx 4$ AU in Figure \ref{fig:e_var_quad}. At these locations, the relative velocities of colliding objects naturally become very small, promoting their agglomeration (rather than fragmentation) and growth.


\section{Self-consistent Models of Self-gravitating, Rigidly Precessing Disks}
\label{sect:modes}


We now use our results to assess the possibility of constructing self-consistent models of long-lived, self-gravitating eccentric disks orbiting massive central objects. Such models could describe, for example, the eccentric nuclear stellar disks around supermassive black holes observed in the centers of some galaxies. 

We will assume that the surface density distribution in the disk is given by equation (\ref{eq:Sigma}), which essentially implies that for each semimajor axis, there is a single, unique value of the disk eccentricity $e_d(a)$, and that eccentric orbits of particles at all semimajor axes have the same orientation $\varpi_d$. This orientation cannot be fixed in time, as the disk's own non-Newtonian potential causes the orbits of individual particles to precess. In other words, $\varpi_d=\varpi_d(t)$. Given the distribution of the surface density $\Sigma_d(a)$ at the pericenter, the question we ask is whether one can determine the profile of $e_d(a)$ that the disk must have to precess coherently as a solid body at a constant rate $\dot\varpi_d$ (i.e. $\varpi_d(t)=\dot\varpi_d t$). This arrangement, obviously, requires $\dot\varpi_d$ to be independent of $a$, since otherwise, differential precession would lead to disk twisting (apsidal misalignment of different parts of the disk),  destroying its coherence. 

Introducing for convenience the complex eccentricity $E_p=k_p+ih_p=e_p e^{i\varpi_p}$ we can combine equations (\ref{eq:sec_eq}) into a single evolution equation for $E_p$:
\ba  
-i\dot E_p=A_d E_p+B_de^{i\varpi_d}.
\label{eq:E_ev}
\ea  
This equation is valid for any object, including the particles or fluid elements comprising the disk and contributing to its potential. Provided that solid-body precession is the only secular effect of the disk self-gravity, i.e. that the disk remains stationary in the frame precessing at the rate $\varpi_d$, we look for solutions with $\dot e_p=0$ (i.e. $e_p(a,t)=e_p(a)$) and $\varpi_p=\varpi_d=\dot\varpi_d t$. Also, by our assumption, at each point the disk eccentricity $e_d$ is the same as the eccentricity of its constituent particles passing through this point, meaning that we need to identify $e_p=e_d$. Plugging the ansatz $E_p=e_d(a)e^{i\dot\varpi_d t}$ into the equation (\ref{eq:E_ev}), one arrives at the following master equation:
\ba  
\left[\dot\varpi_d-A_d(\Sigma_d,a)\right]e_d(a)=B_d(\Sigma_d,e_d,a).
\label{eq:master}
\ea 

This equation represents a self-consistent mathematical framework for determining the radial profile of $e_d$ that an eccentric disk needs to have to be able to precess as a solid body (without changing its shape) under the action of its own self-gravity. The precession rate $\dot\varpi_d$ plays the role of an {\it eigenvalue} of the problem. Equations (\ref{eq:AdBd})-(\ref{eq:Bd_edge}) provide explicit dependencies of $A_d$ and $B_d$ on $\Sigma_d$, $e_d$, and $a$. The dependence is such that (\ref{eq:master}) is an {\it integral equation} for $e_d$. It is linear in $e_d$ and is essentially a Fredholm equation of the second type. Solving this integral equation, we obtain a set of eigenvalues (precession rates $\dot\varpi_d$), as well as the corresponding eigenfunctions (radial profiles of $e_d$; the normalization of $e_d$ remains unconstrained because of the linear nature of equation (\ref{eq:master})). 

This calculation uses the radial distribution of $\Sigma_d$ as an input. As mentioned in \S \ref{sect:method}, when $e_d(a)$ does not change in the course of evolution (or whenever $e_d\ll 1$), the radial profile of the surface density at periastron $\Sigma_d(a)$ remains fixed in the course of secular evolution. 

We defer the detailed exploration of the equation (\ref{eq:master}) for disks with different $\Sigma_d$ profiles to future work. We will simply note here that some of our findings --- divergent behavior of $A_d$ and $B_d$ near the sharp disk edges, the changes of signs of $A_d$ and $B_d$ inside the disk, etc. --- make finding the solutions of this equation  rather nontrivial.


\section{Discussion}
\label{sect:disc}


Our results allow one to efficiently compute the effect of the gravity of an eccentric disk on the secular evolution of astrophysical objects coplanar with the disk. This work provides a natural generalization of the earlier calculation of \citet{SR15b}, in which the secular potential was computed for eccentric disks with $\Sigma_d(a)$ and $e_d(a)$ given by power laws of $a$ only. Even prior to that, \citet{Heppen} and \citet{Ward} derived the disturbing function for a particular case of {\it axisymmetric} power-law disks. Our present results extend these calculations for {\it arbitrary} behaviors of $\Sigma_d(a)$ and $e_d(a)$, allowing a much broader range of applications of our framework. 

The accuracy with which our lowest-order theory works depends on both the behavior of the disk eccentricity and the particle eccentricity. The results of \S \ref{sect:ecc} demonstrate that our secular theory becomes inaccurate when $e_d\gtrsim 0.2$ or so; the actual value of $e_d$ at which this happens depends on both the location in the disk (see Figure \ref{fig:A_diff}) and the radial profile of $e_d$ (see Figure \ref{fig:e_var_quad}). Moreover, our results clearly show that even for nearly circular disks, the behavior of $e_p$ can be singular at certain locations (at least for $\Sigma_d$ profiles considered in this work), resulting in very high $e_p\sim 1$ even at semimajor axes where $e_d(a)\ll 1$. As evidenced by Figure \ref{fig:zoom}, this naturally leads to deviations from our theory at these locations, even for low-$e_d$ disks.

Real astrophysical disks have rather different values of $e_d$. For example, the stellar disk in M31 has a substantial eccentricity, $e_d\sim 0.5$ \citealt{Tremaine95,Brown,Peiris}). Our secular theory is unlikely to provide a good description of the secular dynamics in this system, as even its topology of the phase space should look very different from that corresponding to the disturbing function (\ref{eq:R_gen}). A higher-order extension of our approach, such as that presented recently in \citet{Sefilian} to generalize the \citet{SR15b} calculations to fourth order in $e_p$, may provide a better tool for studying disks with nonnegligible $e_d$.

On the other hand, disk eccentricity can be low enough in gaseous  protoplanetary disks in binary stellar systems. Simulations demonstrate that under certain circumstances (moderately high binary eccentricity), a disk orbiting one of the binary components can have rather low eccentricity, at the level of several percent \citep{Marzari,Regaly,Muller}. The same is true for circumbinary protoplanetary disks on AU scales \citep{Meschiari}. In such systems, our theory should be well suited for describing both the effect of the disk gravity on planetesimal motion and planet formation in binary systems \citep{RS15a,RS15b}, as well as for studying the self-consistent dynamics of the gaseous component of the disk \citep{Ogilvie} driven by its self-gravity, pressure forces, etc.

Eccentric planetary rings typically have $e_d\sim 10^{-3}-10^{-2}$ \citep{Elliot}, which is also low enough for our theory to work well in describing the effect of the ring gravity on the secular motion of the adjacent objects (including the ring particles themselves; see \S \ref{sect:modes}).

However, when applying our framework to real planetary rings, a word of caution is in order. One of the underlying assumptions used in deriving the linearized equation (\ref{eq:Sig_expand}) is that $\zeta e_d=de_d/d\ln a\ll 1$. At the same, time the $\epsilon$ ring of Uranus exhibits a change of eccentricity $\Delta e_d\approx 7.11\times 10^{-4}$ over the width $\Delta a\approx 58.1$ km of the ring with a mean semimajor axis $a=51,149$ km \citep{French}. We can evaluate $\zeta e_d\approx a\Delta e_d/\Delta a\approx 0.6$, which is certainly not small compared to unity. This means that $\Sigma$ exhibits strong variation along the eccentric orbits of ring particles. As a result, both the expansion (\ref{eq:Sig_expand}) and our resultant framework (\ref{eq:R_gen}), (\ref{eq:AdBd})-(\ref{eq:Bd_edge}) may become inaccurate when applied to rings with $\zeta e_d\sim 1$. A similar issue was previously discussed in \S \ref{sect:ecc_prof}.  

The cost involved in calculating secular evolution according to our approach is relatively low --- it requires computation of only one-dimensional integrals involving $\Sigma_d(a)$ and $e_d(a)$ (to obtain coefficients $A_d$ and $B_d$). This is to be contrasted with the direct approach to computing secular potential, embodied by equation (\ref{mainint}), in which one first needs to carry out the two-dimensional integration over the full disk to obtain the potential at every point and then one additional integration to average it over the particle trajectory. Our procedure is clearly less numerically intensive and reproduces direct calculations very well in the low-$e$ limit. This allowed us to use it for exploring different characteristics of secular motion in the disk potential, which we did in \S \ref{sect:edges}-\ref{sect:modes}. 

The secular dynamics of self-gravitating disks is often explored by modeling the them as collections of narrow adjacent rings coupled via the {\it softened} secular gravitational potential \citep{Tremaine2001,Hahn,Touma}. In this approach, the secular potential is approximated by the modified version of the classical Laplace-Lagrange disturbing function \citep{MurrayDermott}, which is regularized via softening to avoid the singularity that arises when the semimajor axes of the rings overlap. This procedure inevitably introduces an ad hoc {\it softening parameter} into the calculation, which inevitably leads to ambiguity of the results, since a physical justification for a particular choice of softening is not obvious.   

Our approach, ascending to the framework developed in \citet{Heppen}, \citet{Ward}, and \citet{SR15b}, does not suffer from this ambiguity. Even though the integrand of the expressions (\ref{eq:Ad_bulk}), (\ref{eq:Bd_bulk}) for $A_d$ and $B_d$ contains Laplace coefficients $b^{(0)}_{1/2}(\alpha)$ and $b^{(1)}_{1/2}(\alpha)$, which diverge as $\alpha\to 1$ (i.e. for particles with semimajor axes inside the radial extent of the disk), their singularity is weak (logarithmic in $\alpha$). As a result, the integrals in equations (\ref{eq:Ad_bulk}), (\ref{eq:Bd_bulk}) are fully convergent {\it without} introducing any ad hoc softening. This makes our calculation of $R_d$ more robust and self-consistent compared to some other treatments. The only possible divergences that remain in our case may arise at the sharp edges of the disk and are truly physical in their nature (see \S \ref{sect:edges}).

Our calculation of the disturbing function explicitly assumes the disk to have zero thickness and to be coplanar with the particle orbit. It is easy to show that our results would not change qualitatively if the disk were to have a small vertical thickness $h$, as long as $h$ is small compared to the semimajor axis $a$ in the disk. Quantitative corrections to our results due to the nonzero $h$ should be of order $O(h/a)$ at most. The only substantial difference will arise close to the edges of the disks with sharply truncated $\Sigma_d$, see \S \ref{sect:edges}.

We also explicitly assumed the surface density of the disk $\Sigma$ to have a specific form (\ref{eq:Sigma}), valid when the mass elements comprising the disk move around the central mass on eccentric orbits, which are apsidally aligned and have a unique value of eccentricity for a given semimajor axis. This assumption is natural for fluid disks, in which orbit crossing is impossible in the absence of shocks, as well as for highly collisional planetary rings. However, particulate disks of low optical depth (e.g. nuclear stellar disks) may have a more complicated form of $\Sigma$ because orbit crossings are possible in such systems. We can naturally extend our approach to treat such cases by using the additive nature of gravity: if a more general disk structure can be represented as a superposition of multiple (sub)disks each with $\Sigma$ in the form  (\ref{eq:Sigma}) but different $\Sigma_d(a)$, $e_d(a)$ and apsidal orientation, then the resultant disturbing function $R_d$ will be a sum of individual contributions in the form (\ref{eq:R_gen}) produced by each of the subdisks.


\section{Summary}  
\label{sect:sum}


We explored the secular effect of a massive, razor-thin, eccentric, apsidally aligned disk on the motion of coplanar objects in the combined potential of such a disk (considered as a perturbation) and a central point mass. This problem is of great importance for many astrophysical disks (both gaseous and particulate), ranging from planetary rings to nuclear stellar disks in centers of galaxies harboring supermassive black holes. Our main findings are briefly summarized below. 

\begin{itemize}

\item We developed a general analytical framework for computing the secular disturbing function due to the gravity of an aligned eccentric disk. This disturbing function has the conventional Laplace-Lagrange form with coefficients that contain one-dimensional integrals over the semimajor axis only and do not involve softening of any form. It is valid for arbitrary radial profiles of the disk surface density $\Sigma_d$ and eccentricity $e_d$, and works for coplanar objects orbiting both inside and outside the disk (for disks with edges). 

\item We verified the accuracy of our analytical calculation using direct orbit integrations, finding excellent agreement in the limit of low eccentricity (both disk and particle) for all radial profiles of $\Sigma_d$ and $e_d$ that we considered. Our framework is accurate at the $\lesssim 10\%$ level for disk eccentricities $e_d\lesssim 0.1-0.2$. However, these figures strongly depend on both the location in the disk and the disk model, e.g. eccentricity profile $e_d(a)$.

\item Our calculations demonstrate that free precession of particle orbits in the potential of a smooth eccentric disk can naturally change from prograde to retrograde, and vice versa. Thus, the retrograde free precession previously found for certain types of disks (with a power-law dependence of density on radius) does not hold in general. At the locations where the free precession rate changes sign, forced particle eccentricities reach very high values (formally diverge). 

\item Sharp features in the disk surface density distribution, such as disk edges or gaps, inevitably  change the sense of free precession (e.g. prograde in the gaps, retrograde in the disk). In disks with sharp edges (where $\Sigma_d$ discontinuously drops to zero), the free precession rate formally diverges at the edge.  

\item Using our results, we formulated a general framework for computing the eccentricity profile of a disk (with a prescribed surface density profile  $\Sigma_d(a)$), which is required for it to be precessing rigidly due to its self-gravity alone, while maintaining stationary structure in the frame of precession.
    
\end{itemize}

In the future, we plan to extend our approach for understanding the secular dynamics of {\it twisted} (i.e. {\it apsidally misaligned}) eccentric disks (Davydenkova \& Rafikov, in prep.), as well as for exploring the inclination dynamics in the case of a non-coplanar (warped) disk. Exploration of the modal solutions (i.e. long-lived, rigidly precessing disk models discussed in \S \ref{sect:modes}) based on our secular framework is another logical application of our results.

\acknowledgements

We are grateful to Ryan Miranda, Antranik Sefilian, Kedron Silsbee and Scott Tremaine for useful discussions. Financial support for this study has been provided by the NSF via grant AST-1409524 and NASA via grant 15-XRP15-2-0139.


\bibliographystyle{apj}
\bibliography{references}


\appendix


\section{Calculation of the secular potential of the disk}
\label{sect:disk_potential}


In this section, we present a calculation of the disturbing function $R_d$ due to an eccentric disk with a surface density in the form (\ref{eq:Sigma}) in the low-eccentricity limit. We do not assume any specific forms for the functions $\Sigma_d(a)$ and disk eccentricity $e_d(a)$ apart from requiring them to be twice differentiable. Following the recipe presented in \citet{SR15a}, we first write the disturbing function $R_d$ due to the disk $\mathbb{D}$ as
\ba  
R_d=R(\mathbb{D})=-\left \langle \Phi_d(r_d,\phi_d)\right \rangle,~~~\Phi_d(r_d,\phi_d) = - G \int_{\mathbb{D}}  \frac{\Sigma(r_d,\phi_d) r_d d r_d d\theta}{\sqrt{r_p^2+r_d^2-2r_pr_d\cos{\theta}}},
\label{mainint}
\ea  
where brackets $\langle ...\rangle$ represent time averaging over planetesimal orbital motion, ${\bf r}_p$ is the instantaneous planetesimal radius vector ($r_p=|{\bf r}_p|$, making an angle $\phi_p$ with the planetesimal apsidal line), ${\bf r}_d$ is the radius vector of a surface element of the disk ($r_d=|{\bf r}_d|$, making an angle $\phi_d$ with the disk apsidal line), and $\theta=\phi_d+\varpi_d-\phi_p-\varpi_p$ is the angle between the two; see Figure \ref{fig:geometry}. 

Note that the indirect potential vanishes identically for a disk composed of mass elements moving on purely Keplerian orbits around the central mass. This is easy to understand based on Kepler's second law, which in our case means that the mass elements on opposite sides of a given elliptical trajectory (with respect to its focus) exert equal and opposite forces on the central mass. As a result, mass elements with orbits within a given semimajor axis interval $da$ exert no net force on the central mass and do not cause its reflex motion, meaning that the indirect potential is zero.

We need to evaluate the integral (\ref{mainint}) to second order in eccentricity, i.e. keeping the terms $O(e_p^2)$ and $O(e_pe_d)$. At the same time, we do not need to keep the terms not involving $e_p$ (e.g. $O(e_d)$ or $O(e_d^2)$), as those will vanish upon substitution into the Lagrange equations \citep{MurrayDermott}. Nor do we need to keep terms of higher order in $e_d$ than those listed above. Given that some of the variables entering the expression (\ref{eq:Sigma}) are functions of $a_p$ and not $r_p$, we need to express them through $r_p$ via $a_p\approx r_p(1+e_d(r_p)\cos\phi_d)+O(e_d^2)$. With this in mind, we can expand the expression (\ref{eq:Sigma}) to lowest order in $e_d\ll 1$ and $de_d/d\ln r_d\ll 1$ as \citep{SR15a}
\ba
\Sigma(r_d,\phi_d)\approx \Sigma_d(r_d)+r_d\fr{d}{dr_d}\left[\Sigma_d(r_d)e_d(r_d)\right]\cos\phi_d-\Sigma_d(r_d)e_d(r_d).
\label{eq:Sig_expand}
\ea
As will become obvious later, the integral (\ref{mainint}) over the last term in equation (\ref{eq:Sig_expand}) does not contribute to $R_d$ at the required level of accuracy, so we drop it from the consideration from now on.

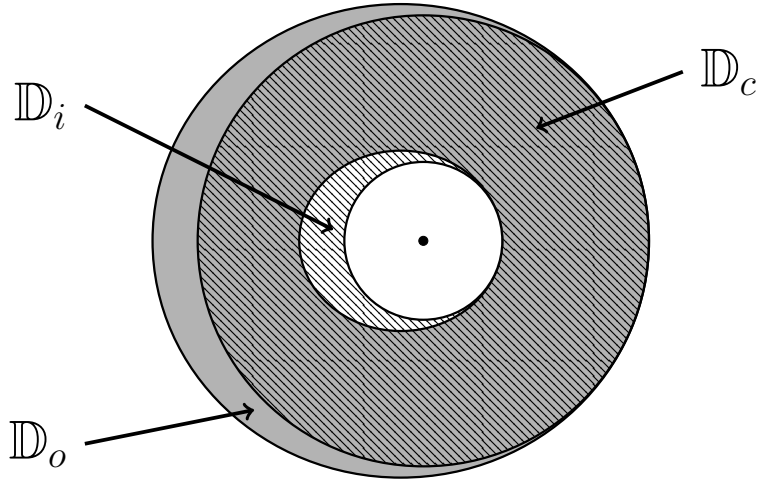
\begin{figure}[h]
\begin{center}
\begin{tikzpicture}[scale=1.5]
 \filldraw[color=gray!60](-0.2,0) ellipse (2.2 cm and 2.1 cm);
 \draw[line width = 0.3 mm](-0.2,0) ellipse (2.2 cm and 2.1 cm);
 \filldraw[color=white](-0.2,0) ellipse (0.9 cm and 0.8 cm);
 \draw[line width =0.3mm](-0.2,0) ellipse (0.9 cm and 0.8 cm);
 \draw[pattern=north west lines, line width =0.3 mm](0,0) circle[radius=2cm];
 \filldraw[color=white](0,0) circle[radius=0.7cm];
 \filldraw[color=black](0,0) circle[radius=0.04cm];
 \draw[line width =0.3 mm](0,0) circle[radius=0.7cm];
 \draw[<-, line width=0.5mm] (1,1) -- (2.3,1.5);
 \node[label={right: \huge$\mathbb{D}_c$}] () at (2.3,1.5){};
 \draw[->, line width=0.5mm] (-3,-1.8) -- (-1.5,-1.5);
 \node[label={left: \huge$\mathbb{D}_o$}] () at (-3,-1.8){};
 \draw[->, line width=0.5mm] (-3,1.2) -- (-0.8,0.1);
 \node[label={left: \huge$\mathbb{D}_i$}] () at (-3,1.2){};
\end{tikzpicture}
\end{center}
\caption{Decomposition of the disk into three distinct regions ($\mathbb{D}_c$, $\mathbb{D}_i$, $\mathbb{D}_o$) over which potential is integrated. The gray area corresponds to the integration area, that is, the physical disk $\mathbb{D}=\mathbb{D}_o+\mathbb{D}_c-\mathbb{D}_i$. The hatched region $\mathbb{D}_c$ is circular. See text for details.}
\label{disk}
\end{figure}

Following \citep{SR15a}, we split the expression (\ref{mainint}) into three integrals over the regions $\mathbb{D}_c$, $\mathbb{D}_i$ and $\mathbb{D}_o$, as shown in Figure \ref{disk}. We define $\mathbb{D}_c$ to be a circular region with an inner radius equal to the periastron distance of the inner edge of the disk, $r_{d,{\rm in}}=a_1(1-e_d(a_1))$, and the outer radius equal to the periastron distance of the outer edge of the disk $r_{d,{\rm out}}=a_2(1-e_d(a_2))$. Region $\mathbb{D}_i$ is the {\it crescent} region confined between the inner circle of $\mathbb{D}_c$ and the inner ellipse of the integration region $\mathbb{D}$; $\mathbb{D}_o$ is the {\it crescent} region confined between the outer circle of $\mathbb{D}_c$ and the outer ellipse of the integration region $\mathbb{D}$. In other words,
\ba    
\mathbb{D}=\mathbb{D}_c-\mathbb{D}_i+\mathbb{D}_o, ~~~\mbox{and}~~~
R(\mathbb{D})=R(\mathbb{D}_c)-R(\mathbb{D}_i)+R(\mathbb{D}_o).
\label{eq:split}
\ea   

We now consider the three contributions to $R(\mathbb{D})$ separately.


\subsection{Calculation of the part of the disturbing function due to the circular region $\mathbb{D}_c$}


We start by computing the integral over the circular annulus $\mathbb{D}_c$. When using the approximation (\ref{eq:Sig_expand}), it is natural to divide $R(\mathbb{D}_c)$ into two parts: one containing $\cos\phi_d$ (we call it $I_1$) and the one independent of $\phi_d$ ($I_2$). We evaluate the $\phi_d$-independent contribution first:
\ba  
I_1 =   \left \langle G \int\limits_{r_{d,{\rm in}}}^{r_{d,{\rm out}}} dr_d~ \Sigma_d(r_d) r_d \int\limits_{0}^{2\pi}  \frac{d\theta}{\sqrt{r_p^2+r_d^2-2r_pr_d\cos{\theta}}} \right \rangle,
\ea   
This integral has a singularity at $r_p=r_d$, so we break it into two parts, $r_p<r_d$ and $r_p>r_d$, while also rewriting the inner integral as a Laplace coefficient $b_{1/2}^{(0)}$, see definition (\ref{eq:Laplace}):
\ba  
I_1  =  \pi G \left \langle \int\limits_{r_{d,{\rm in}}}^{r_p} \Sigma_d(r_d) \frac{r_d}{r_p} b_{1/2}^{(0)}\left(\frac{r_d}{r_p}\right)dr_d+ 
\int\limits_{r_p}^{r_{d,{\rm out}}} \Sigma_d(r_d) b_{1/2}^{(0)}\left(\frac{r_p}{r_d}\right)dr_d \right\rangle.
\ea  
Now let us make a change of variables in both integrals so as to get rid of the time dependence in the  Laplace coefficients, entering via $r_p$. In the first integral, we set $\alpha=r_d/r_p<1$, while in the second, we use $\alpha=r_p/r_d<1$. We will also approximate $r_{d,{\rm in}}\approx a_1$ and $r_{d,{\rm out}}\approx a_2$, which, as we will see, introduces error of negligible order. Now we have
\ba  
I_1  =I_1^{(0)}+I_1^{(1)} = \pi G \left \langle \int\limits_{{a_1}/r_p}^{1} r_p \Sigma_d(\alpha r_p) \alpha b_{1/2}^{(0)}(\alpha)d\alpha+ 
\int\limits^{1}_{r_p/a_2} r_p\Sigma_d(r_p/\alpha) \alpha^{-2}  b_{1/2}^{(0)}(\alpha)d\alpha \right\rangle.
\ea   

Our goal is to get rid of triangular brackets, that is to time average the integrals. The time dependence is hidden in $r_p$ and has the form $r_p(t)=a_p(1-e_p\cos E(t))$ where $e_p\ll 1$ and $E$ is the eccentric anomaly of the particle orbit. Since we are only interested in terms which are no more than quadratic in eccentricity (of both particle and the disk), we next expand the integrals in a Taylor series over $r_p-a_p=-a_p e_p \cos E(t)$ up to second order and average over the planetesimal orbit using the relations $\langle \cos E\rangle =-e_p/2$, $\langle \cos^2 E\rangle =1/2$. Note that the dependence on $r_p$ is also present in the limits of integration, which, upon this Taylor expansion, give rise to the {\it nonintegral} boundary terms. We omit the tedious process of writing out the Taylor expansion and subsequent averaging and present just the result:
\ba  
I_1^{(0)} &\approx & \pi G a_p \fr{e^2_p}{2}\int\limits^{1}_{a_1/a_p}\left.\left[\fr{d}{dr_p}(\alpha r_p\Sigma_d(\alpha r_p)) + \fr{a_p}{2}\fr{d^2}{dr_p} (\alpha r_p\Sigma_d(\alpha r_p))\right]\right|_{r_p=a_p} b_{1/2}^{(0)}(\alpha)d\alpha +
\nonumber\\
&+&\pi G \fr{e_p^2}{4}\left[-\fr{a_1^3}{a^2_p}\Sigma_d(a_1) b_{1/2}^{(0)\prime}\left(\fr{a_1}{a_p}\right)+\fr{a_1^3}{a_p}\Sigma_d^\prime(a_1)b_{1/2}^{(0)}\left(\fr{a_1}{a_p}\right)+\fr{a_1^2}{a_p}\Sigma_d(a_1) b_{1/2}^{(0)}\left(\fr{a_1}{a_p}\right) \right],
\label{eq:I10}
\ea  
where we dropped the terms independent of $e_p$  --- such terms vanish when substituted into the Lagrange evolution equations. For $I_1^{(1)}$ we similarly have 
\ba  
I_1^{(1)} &\approx &
\pi G a_p^2 \fr{e^2_p}{2}
\int\limits^{1}_{a_p/a_2}\left.\left[\frac{1}{r_p}\frac{d}{dr_p}\left(r_p\Sigma_d\left(\alpha^{-1}r_p\right)\right)+\frac{1}{2}\frac{d^2}{dr_p^2}\left(r_p\Sigma_d\left(\alpha^{-1}r_p\right)\right)\right]\right|_{r_p=a_p}
\alpha^{-2}b_{1/2}^{(0)}(\alpha)d\alpha 
\nonumber\\
&-& \pi G \fr{e_p^2}{4}\left[
2a_2\Sigma_d(a_2)b_{1/2}^{(0)}\left(\fr{a_p}{a_2}\right) 
+a_p\Sigma_d(a_2)b_{1/2}^{(0)\prime}\left(\fr{a_p}{a_2}\right)
+a_2^2\Sigma_d^{\prime}(a_2)b_{1/2}^{(0)}\left(\fr{a_p}{a_2}\right)\right].
\label{eq:I11}
\ea

Now we turn to calculation of the integral contribution $I_2$ containing $\phi_d=\theta+v$, where we defined $v=\phi_p+\varpi_p-\varpi_d$. Let us introduce a new auxiliary function, $Q(r_d)=d\left[e(r_d)\Sigma_d(r_d)\right]/dr_d.$ Then,
\ba   
I_2  = G\left \langle \int\limits_{r_{d,{\rm in}}}^{r_{d,{\rm out}}} Q(r_d) r^{2}_d   \int\limits_{0}^{2\pi} \frac{\cos\theta \cos v d\theta}{\sqrt{r_p^2+r_d^2-2r_pr_d\cos{\theta}}}dr_d \right \rangle,
\ea  
where we expanded $\cos{(\theta+v)}$ and took into account that the term proportional to $\sin\theta$ evaluates to zero. 

Next, we divide the integral over $dr_d$ into two parts, one for $r_d<r_p$ and another for $r_d>r_p$, just like it was done for $I_1$. 
We again approximate $r_{d,{\rm in}}\approx a_1$ and $r_{d,{\rm out}}\approx a_2$, as the linear corrections to these expressions would lead to a contribution, which is third order in eccentricity.
We then, again, make a change of variables in both resulting integrals, setting $\alpha=r_d/r_p<1$ in the first one and $\alpha=r_p/r_d<1$ in the second. As a result, using the definition of the Laplace coefficient $b_{1/2}^{(1)}$, we obtain the following expression:
\ba  
I_2  =I_2^{(0)}+I_2^{(1)} = \pi G  \left \langle \int\limits_{{a_1}/r_p}^{1} \cos v~r^2_p Q(\alpha r_p) \alpha^2 b_{1/2}^{(1)}(\alpha)d\alpha+ 
\int\limits^{1}_{r_p/a_2} \cos v ~r_p^2 Q(\alpha^{-1}r_p) \alpha^{-3}  b_{1/2}^{(1)}(\alpha)d\alpha \right\rangle.
\ea  

There are two main differences in our subsequent expansion over $r_p-a_p$ compared to the previous case. First, $Q(r_d)\sim O(e_d)$ and, thus, in the expansion, we only need to retain terms up to first order in $e_p$. Second, now we have the time dependence of the integrand also through $\cos v(t) =\cos(\Delta\varpi + \phi_p(t))=\cos(\Delta\varpi + E(t)+e_p\sin E(t))$, where we defined $\Delta\varpi=\varpi_p-\varpi_d$ and used the relation $\phi_p=E+e_p\sin E$. The easiest way to deal with the calculation is to consider $r_p$ as a function of $e_p$, $r_p=a_p(1-e_p\cos E),$ and to expand the integrals directly over $e_p$. We start with $I_2^{(0)}$:
\ba   
I_2^{(0)}
 =\pi G \left \langle \int\limits_{{a_1}/(a_p(1-e_p\cos E))}^{1} \cos(\Delta\varpi + E+e_p\sin E)r_p^2(e_p,E)Q(\alpha r_p(e_p,E)) \alpha^2 b_{1/2}^{(1)}(\alpha)d\alpha \right\rangle.
\ea   
We need the zeroth and first-order (in $e_p\ll 1$) terms of the Taylor expansion of this expression at $e_p=0.$ We will omit the technicalities of writing out the expansion and present the result:
\ba   
I_2^{(0)}&\approx & \pi G  \int\limits_{{a_1}/a_p}^{1} a_p^2 Q(\alpha a_p) \alpha^2 b_{1/2}^{(1)}(\alpha)d\alpha\left \langle\cos(\Delta\varpi + E)\right\rangle 
- \pi G e_p  \fr{a_1^3}{a_p} b_{1/2}^{(1)}\left(\fr{a_1}{a_p}\right)Q(a_1) \left \langle\cos(\Delta\varpi + E)\cos(E)\right\rangle
\\ 
&-& \pi G e_p  \int\limits_{{a_1}/a_p}^{1} \left(a_p^2 Q(\alpha a_p)  \left \langle\sin(\Delta\varpi + E)\sin(E)\right\rangle 
+ a_p\left.\fr{d}{dr_p}\left(r_p^2Q(\alpha r_p)\right)\right|_{r_p=a_p} \left \langle\cos(\Delta\varpi + E)\cos(E)\right\rangle\right)
\alpha^2 b_{1/2}^{(1)}(\alpha)d\alpha  .
\nonumber
\ea   
Using the averages of the trigonometric functions of $E$,
\ba   
\label{av1}
\left \langle\cos(\Delta\varpi + E)\right\rangle 
= -\fr{e_p}{2}\cos\Delta\varpi,~~~
\left \langle\sin(\Delta\varpi + E)\sin(E)\right\rangle
= \left \langle\cos(\Delta\varpi + E)\cos(E)\right\rangle
=\fr{1}{2}\cos\Delta\varpi,
\ea  
based on $\langle\sin E\rangle=0$, $\langle\sin E\cos E\rangle=0$, and  $\langle\sin^2 E\rangle=1/2$, the final result for $I^{(0)}_2$ becomes
\ba  
 I^{(0)}_2 \approx - \pi G \fr{e_p}{2}\cos\Delta\varpi \left[\int\limits_{{a_1}/a_p}^{1} \left(4 Q(\alpha a_p)+  a_p\left.\fr{d}{dr_p}\left(Q(\alpha r_p)\right)\right|_{r_p=a_p}\right)a_p^2 \alpha^2 b_{1/2}^{(1)}(\alpha)d\alpha +   a_1^2b_{1/2}^{(1)}\left(\fr{{a_p}}{a_1}\right)Q(a_1) \right].
 \label{eq:I20}
\ea  

Following a similar procedure, for $I_2^{(1)}$ we have 
\ba  
I_2^{(1)}\cong - \pi G \fr{e_p}{2}\cos\Delta\varpi \left[\int\limits_{{a_p}/a_2}^{1}\left( 4 Q\left(\fr{a_p}{\alpha}\right)   +  a_p  \left.\fr{d}{dr_p}\left(Q\left(\fr{r_p}{\alpha}\right)\right)\right|_{r_p=a_p}\right)\fr{a_p^2}{\alpha^{3}} b_{1/2}^{(1)}(\alpha)d\alpha  - a_2^2 b_{1/2}^{(1)}\left(\fr{a_p}{a_2}\right)Q(a_2) \right].
\label{eq:I21}
\ea  

Equations (\ref{eq:I10}), (\ref{eq:I11}), (\ref{eq:I20}), and (\ref{eq:I21}) provide the desired contribution to the disturbing function from the circular annulus $\mathbb{D}_c$ since $R(\mathbb{D}_c)=I_1^{(0)}+I_1^{(1)}+I_2^{(0)}+I_2^{(1)}$.


\subsection{Calculation of the Part of the Disturbing Function Due to $\mathbb{D}_i$}


Now let us turn to the inner crescent $\mathbb{D}_i$. As the eccentricity of our disk is small, the width of this crescent is already $O(e_d)$, thus, we only need to keep the terms up to first order in $e_p$ in our expansion. Following the recipe in \citet{SR15a}, we write
\begin{equation}
R(\mathbb{D}_i) = -\left \langle G  \int\limits_{a_1(1-e_1)}^{a_1(1+e_1)} \frac{r_d}{r_p}\Sigma(r_d) dr_d \int\limits_{\xi-\Delta\varpi - \phi_p}^{2\pi-\xi-\Delta\varpi - \phi_p} \frac{d\theta}{\sqrt{1+ \alpha^{2} - 2\alpha \cos\theta}} \right \rangle,
\label{eq:RDi}
\end{equation}
where $\xi(r_d)  = \arccos ((a_1-r_d)/e_1a_1)$ determines the azimuthal extent of the crescent for a given $r_d$ \citep{SR15a}, and $\alpha = r_d/r_p$.

For the function in the inner integral, we can use the Fourier expansion,
\ba  
\left(1 + \alpha^{2} - 2\alpha \cos{\theta}\right)^{-1/2} = \frac{1}{2} b_{1/2}^{(0)}(\alpha) + \sum_{j = 1}^\infty b_{1/2}^{(j)}(\alpha) \cos{(j \theta)},
\ea  
letting us to calculate the inner integral as
\ba   
&& \int\limits_{\xi-\Delta\varpi - \phi_p}^{2\pi-\xi-\Delta\varpi - \phi_p}\left(\frac{1}{2} b_{1/2}^{(0)}(\alpha) + \sum_{j = 1}^\infty b_{1/2}^{(j)}(\alpha) \cos{(j \theta)}\right)d\theta
 =\left(\pi - \xi(r_d)\right)b_{1/2}^{(0)}(\alpha)+\left.\sum_{j = 1}^\infty \frac{1}{j} b_{1/2}^{(j)}(\alpha) \sin(j\theta)\right|_{\xi-\Delta\varpi - \phi_p}^{2\pi-\xi-\Delta\varpi - \phi_p} 
\nonumber\\
&& = \left(\pi - \xi(r_d)\right)b_{1/2}^{(0)}(\alpha)- \sum_{j = 1}^\infty \frac{2}{j} b_{1/2}^{(j)}(\alpha) \sin\left(j\xi(r_d)\right)\cos\left(j(\Delta\varpi + \phi_p)\right).
 \label{eq:st1}
\ea    

The next step is to remember that we only need to keep terms up to $O(e_d)$ in equation (\ref{eq:RDi}) and to notice that the interval over which radial integration is performed is already $O(e_d)$. As a result, in (\ref{eq:RDi}) we can set $r_d\approx a_1$, $\Sigma_d(r_d)\approx \Sigma_d(a_1)$ and take them out of the integral. We can also set $\alpha\approx a_1/r_p$ in equation (\ref{eq:st1}) for the same reason. As a result, we find
\ba   
R(\mathbb{D}_i) &\approx & - G\Sigma(a_1) \left \langle \frac{a_1}{r_p} \int\limits_{a_1(1-e_1)}^{a_1(1+e_1)}  \left[\left(\pi - \xi(r_d)\right)b_{1/2}^{(0)}\left(\fr{a_1}{r_p}\right)- \sum_{j = 1}^\infty \frac{2}{j} b_{1/2}^{(j)}\left(\fr{a_1}{r_p}\right) \sin\left(j\xi(r_d)\right)\cos\left(j(\Delta\varpi + \phi_p)\right) \right]dr_d\right\rangle
\nonumber\\
&=& -\pi G e_1a_1^{2}\Sigma(a_1) \left \langle r_p^{-1}b_{1/2}^{(0)}\left(\fr{a_1}{r_p}\right)-r_p^{-1}b_{1/2}^{(1)}\left(\fr{a_1}{r_p}\right)\cos\left(\Delta\varpi + \phi_p\right)\right\rangle, 
\label{rsi}
\ea    
where we have used the fact that
\ba   
 \int\limits_{a_1(1-e_1)}^{a_1(1+e_1)}\sin\left(j\xi(r_d)\right)dr_d=e_1a_1\int^{\pi}_{0}\sin(j\xi)\sin(\xi)d\xi=\delta_{j1}\fr{\pi}{2}e_1a_1,
 \ea  
where $\delta_{ij}$ is the Kronecker delta.

Now we just need to expand (\ref{rsi}) up to first order in $e_p$ and then orbit average the resultant expression, which is straightforward using the expressions (\ref{av1}):
\ba   
&& \left \langle r_p^{-1}\left[b_{1/2}^{(0)}\left(\fr{a_1}{r_p}\right)-b_{1/2}^{(1)}\left(\fr{a_1}{r_p}\right)\cos\left(\Delta\varpi + \phi_p\right)\right]\right\rangle 
\approx  a_p^{-1}\left\langle b_{1/2}^{(0)}\left(\fr{a_1}{a_p}\right)-b_{1/2}^{(1)}\left(\fr{a_1}{a_p}\right)\cos(\Delta\varpi + E)\right\rangle 
\nonumber\\
&& +a_pe_p\left\langle\cos E~\left[a_p^{-2}b_{1/2}^{(0)}\left(\fr{a_1}{a_p}\right) + a_p^{-1}a_1a_p^{-2}b_{1/2}^{(0)\prime}\left(\fr{a_1}{a_p}\right)\right] \right\rangle - \left \langle a_pe_p\cos E~ a_p^{-2}b_{1/2}^{(1)}\left(\fr{a_1}{a_p}\right)\cos(\Delta\varpi + E)\right.
\nonumber\\
&&  \left. + a_pe_p\cos E~ a_p^{-1}a_1a_p^{-2}b_{1/2}^{(1)\prime}\left(\fr{a_1}{a_p}\right)\cos(\Delta\varpi + E) - e_pa_p^{-1}b_{1/2}^{(1)}\left(\fr{a_1}{a_p}\right)\sin E~ \sin(\Delta\varpi + E)\right\rangle 
\nonumber\\ 
&& 
=a_p^{-1}b_{1/2}^{(0)}\left(\fr{a_1}{a_p}\right)+ \fr{e_p}{2} a_p^{-1}\left[b_{1/2}^{(1)}\left(\fr{a_1}{a_p}\right) - \fr{a_1}{a_p}b_{1/2}^{(1)\prime}\left(\fr{a_1}{a_p}\right)\right]\cos\Delta\varpi.
\ea  
As a result, by dropping the first term not containing $e_p$, one finally finds
\ba    
R(\mathbb{D}_i) \approx 
- \pi G \Sigma(a_1)e_1\fr{e_p}{2}\fr{a_1^{2}}{a_p}\left[b_{1/2}^{(1)}\left(\fr{a_1}{a_p}\right) - \fr{a_1}{a_p}b_{1/2}^{(1)\prime}\left(\fr{a_1}{a_p}\right)\right]\cos\Delta\varpi.
\label{eq:R_Di}
\ea


\subsection{Calculation of the Part of the Disturbing Function Due to $\mathbb{D}_o$}


This case has only minor differences from the previous one, so we will omit the details:
\ba  
&& R(\mathbb{D}_o) = \left \langle G  \int\limits_{a_2(1-e_2)}^{a_2(1+e_2)} \Sigma(r_d) dr_d \int\limits_{\xi-\Delta\varpi - \phi_p}^{2\pi-\xi-\Delta\varpi - \phi_p} \frac{d\theta}{\sqrt{1+ \alpha^{2} - 2\alpha \cos\theta}} \right \rangle 
\nonumber\\ 
&& \approx\pi G e_2a_2\Sigma(a_2) \left \langle b_{1/2}^{(0)}\left(\fr{r_p}{a_2}\right)- b_{1/2}^{(1)}\left(\fr{r_p}{a_2}\right)\cos\left(\Delta\varpi + \phi_p\right)\right\rangle
\nonumber\\ 
&& \approx \pi G e_2a_2\Sigma(a_2) \left[b_{1/2}^{(0)}\left(\fr{a_p}{a_2}\right) + e_pb_{1/2}^{(1)}\left(\fr{a_p}{a_2}\right)\cos\Delta\varpi +\fr{e_p}{2}a_pa_2^{-1}b_{1/2}^{(1)\prime}\left(\fr{a_p}{a_2}\right)\cos\Delta\varpi\right]. 
\ea  
Thus, again dropping the first, $e_p$-independent term, we eventually arrive at
\be
R(\mathbb{D}_o)\approx 
\pi G \Sigma(a_2) e_2\fr{e_p}{2}a_2\left[2b_{1/2}^{(1)}\left(\fr{a_p}{a_2}\right)+\fr{a_p}{a_2}b_{1/2}^{(1)\prime}\left(\fr{a_p}{a_2}\right)\right]\cos\Delta\varpi.
\label{eq:R_Do}
\ee


\subsection{The full result}


The full expression resulting from evaluating the integral (\ref{mainint}) is given by $R_d=I_1^{(0)}+I_1^{(1)}+I_2^{(0)}+I_2^{(1)}+R(\mathbb{D}_i)+R(\mathbb{D}_o)$ and can be written out explicitly using equations (\ref{eq:I10}), (\ref{eq:I11}), (\ref{eq:I20}), (\ref{eq:I21}), (\ref{eq:R_Di}), and  (\ref{eq:R_Do}). When doing this, we introduce two important modifications.

First, it is possible to combine the different terms corresponding to the outer and inner disks into a single expression using certain properties of the Laplace coefficients. For example, we can combine $I_1^{(0)}$ and $I_1^{(1)}$ by changing the variable $\alpha\to \alpha^{-1}$ in equation (\ref{eq:I11}) for $I_1^{(1)}$ and then manipulate the result using the fact that $b_s^{(j)}(\alpha^{-1})=\alpha^{2s} b_s^{(j)}(\alpha)$. This procedure turns $I_1^{(1)}$ in equation (\ref{eq:I11}) into the expression analogous to equation (\ref{eq:I10}) but with the limits of integration running from $1$ to $a_2/a_p$. This allows us to combine it with $I_1^{(0)}$ to get an expression analogous to equation (\ref{eq:I10}) with integration running from $a_1/a_p<1$ to $a_2/a_p>1$ (for orbits {\it within} the disk, $a_1<a_p<a_2$). The integrand of this final expression has a weak (logarithmic) singularity at $\alpha=1$, but the integral itself is fully convergent. This procedure is then repeated to combine $I_2^{(0)}$ and $I_2^{(1)}$. In addition, differentiating the identity $b_s^{(j)}(\alpha^{-1})=\alpha^{2s} b_s^{(j)}(\alpha)$ we obtain the relation for $b_s^{(j)\prime}(\alpha^{-1})$, which allows us to also merge $R(\mathbb{D}_i)$ and $R(\mathbb{D}_o)$ into a single expression.  

Second, in all these expressions, we switch from $d/dr_p$ to derivatives with respect to the argument (denoted by prime), e.g. $dQ(\alpha r_p)/dr_p=\alpha Q^\prime(a)|_{a=\alpha r_p}$

As a result of performing these steps, one finds that $R_d$ can be written in the form (\ref{eq:R_gen}) with the coefficients $A_d$ and $B_d$ given by the following expressions:
\ba
A_d & = & A_d^{\rm bulk}+A_d^{\rm edge},~~~~
B_d = B_d^{\rm bulk}+B_d^{\rm edge},
\label{eq:AdBd}\\
A_d^{\rm bulk} & = & \frac{\pi G}{2a_p n_p} 
\int \limits^{a_2/a_p}_{a_1/a_p} \alpha b_{1/2}^{(0)}(\alpha) \left(a^2\Sigma_d(a)\right)^{\prime\prime}\Big |_{a=\alpha a_p}d\alpha, 
\label{eq:Ad_bulk}\\
A_d^{\rm edge} & = & -\frac{\pi G}{2a_p n_p} 
\left[\left(\frac{a}{a_p}\right)^2 b_{1/2}^{(0)}\left(\frac{a}{a_p}\right) \left(a\Sigma_d(a)\right)^\prime
-\left(\frac{a}{a_p}\right)^3 b_{1/2}^{(0)\prime}\left(\frac{a}{a_p}\right) \Sigma_d(a)\right]\Bigg |_{a=a_1}^{a=a_2},
\label{eq:Ad_edge}\\
B_d^{\rm bulk} & = & - \frac{\pi G}{2 n_p}  
\int \limits^{a_2/a_p}_{a_1/a_p} \alpha^2 b_{1/2}^{(1)}(\alpha) 
\left[4\left(\Sigma_d(a) e_d(a)\right)^\prime+a\left(\Sigma_d(a) e_d(a)\right)^{\prime\prime}\right]
\Big |_{a=\alpha a_p}d\alpha,
\label{eq:Bd_bulk}\\
B_d^{\rm edge} & = & \frac{\pi G}{2 a_p^3 n_p}
\left[
a^3 b_{1/2}^{(1)}\left(\frac{a}{a_p}\right) \left(\Sigma_d(a)e_d(a)\right)^\prime
+ a^{2} \left(b_{1/2}^{(1)}\left(\fr{a}{a_p}\right) - \frac{a}{a_p}b_{1/2}^{(1)\prime}\left(\fr{a}{a_p}\right)\right)\Sigma_d(a) e_d(a)\right]\Bigg |_{a=a_1}^{a=a_2},
\label{eq:Bd_edge}
\ea 
where $F(a)|_{a=a_1}^{a=a_2}\equiv F(a=a_2)-F(a=a_1)$.
If we consider a disk model with power-law profiles of  $\Sigma_d(a)\propto a^{-p}$ and $e_d(a)\propto a^{-q}$, the results (\ref{eq:AdBd})-(\ref{eq:Bd_edge}) reduce to the corresponding formulae in \citet{SR15a}.


\section{Gap}
\label{sect:Gap}


The ring profile used in \S \ref{sect:edges} to illustrate the effects of sharp disk edges on the free precession rate is given by $\Sigma_d(a)=\Sigma_0(a,\nu)$ with 
\ba
\Sigma_0(a,\nu)=100\times \exp \left[1 -\exp \left(\frac{1}{0.1 +5^6\left(\left(x-3\right)^{\nu}+1\right)^{-6}}\right)\right]~{\rm g ~cm}^{-2},
\label{eq:ring}
\ea
where $x=a/$AU and $\nu$ is a parameter controlling the sharpness of the disk edges; see Figure \ref{fig:edges}.

The presence of a gap of radial width w and relative depth $d<1$ is modeled in \S \ref{sect:gap} using the surface density profile $\Sigma_d(a)=\Sigma_0(a,2)\times {\rm Gap}(a,d,{\rm w})$, where 
\ba   
&&{\rm Gap}(a,d,{\rm w})=1-d \times \exp \left[1 - e^ {u(x,{\rm w})-u(3,{\rm w})}\right],~~~~~u(x,{\rm w})=\left[0.1 +\left(\frac{\textrm{w}}{\left(x-3\right)^2+1}\right)^5\right]^{-1}.
\label{eq:Gap}
\ea

\end{document}